\documentstyle[12pt,epsfig]{article}
\begin{document}

\begin{center}
{\Large \bf {
Two--neutrino double beta decay of $^{76}Ge$ within 
  deformed QRPA: A new suppression mechanism}}

\bigskip 
Fedor \v Simkovic$^{1,2)}$,
Larisa Pacearescu$^{1)}$
and Amand Faessler$^{1)}$

\bigskip

{\it 1)
Institute f\"ur Theoretische Physik  der Universit\"at
T\"ubingen,\\[0pt]
Auf der Morgenstelle 14, D-72076 T\"ubingen, Germany}\\[0pt]

{\it 2)
Department of Nuclear Physics, Comenius University,\\[0pt]
SK--842 48 Bratislava, Slovakia}\\[0pt]

\end{center}

\bigskip

\begin{abstract}
{ 
The effect of deformation on the two-neutrino double decay
($2\nu\beta\beta$-decay) for ground state transition 
$^{76}Ge \rightarrow {^{76}Se}$ is studied in the framework 
of the deformed QRPA with separable Gamow-Teller residual 
interaction. A new suppression mechanism of the 
$2\nu\beta\beta$-decay matrix element based on the difference 
in deformations of the initial and final nuclei is included.
An advantage of this suppression mechanism in comparison with 
that associated with ground state correlations is that it allows
a simultaneous description of the single $\beta$ and the 
$2\nu\beta\beta$-decay.  By performing a detail
calculation of the $2\nu\beta\beta$-decay of $^{76}Ge$,
it is found that the states of  intermediate nucleus lying 
in the region of the Gamow-Teller resonance 
contribute significantly to the matrix element of this process.

}
\end{abstract}

\section{Introduction}
\label{sec: level1}

There is a continuous attempt to improve the accuracy and reliability 
of the calculated single beta ($\beta$) and double beta ($\beta\beta$)
decay nuclear matrix elements . 
Their values affect our understanding of astrophysical processes and
the fundamental properties of neutrino, in particular neutrino mixing
and masses. 

The complexity of the calculation of the double beta decay matrix 
elements consists in the fact that it is a second order process, i.e.,
in addition to the initial and final nuclear states the knowledge
of the complete set of the states of the intermediate nucleus is
required. In solving of this problem different models 
as well as different nuclear structure scenarios were applied
\cite{Fas1,SuCi,eji00}.

Double beta decay can occur in different modes: The neutrinoless 
mode ($0\nu\beta\beta$-decay) 
requires violation of the lepton number and probes new physics
scenarios beyond the Standard Model (SM) of particle physics
\cite{Fas1,SuCi,Doi,klapd,vog02,ver02} . 
The two-neutrino decay ($2\nu\beta\beta$-decay) 
mode is process allowed in the SM, i.e., the half-life of 
this process is free of unknown parameters of the particle
physics side \cite{Fas1,SuCi,eji00,Doi}. 

As the half-lifes of the $2\nu\beta\beta$-decay
have been already measured for about ten 
nuclei, the values of the  associated nuclear matrix element
can be extracted directly. It provides a cross-check on the 
reliability of the matrix element calculations.

Since the nuclei undergoing double beta decay are open shell nuclei
the proton-neutron Quasiparticle Random Phase Approximation
(pn-QRPA) method has been the most employed in the evaluation
of the double beta decay matrix elements \cite{Fas1,SuCi}. In addition,
this approach has succeeded  in reproducing the experimentally 
observed suppression of the $2\nu\beta\beta$-decay transitions
\cite{vogel2,civ1,mut}.
However, a strong sensitivity of the computed matrix elements 
on an increase of the strength of the particle-particle 
residual interaction  in the $1^+$ channel 
leads to a problem of fixing this parameter \cite{Fas1}.
Thus various refinements of the original pn-QRPA have been advanced.

In the last decade the extensions of the pn-QRPA, which include
proton-neutron pairing (full-QRPA) \cite{cheo}, a partial restoration of the
Pauli exclusion principle discarded in the pn-QRPA  (renormalized 
QRPA) \cite{Toi95,simn96}
and higher order RPA corrections \cite{radu91,radu00}, 
have received much attention 
in the context of the nuclear structure calculations for the 
double beta decay. In spite of the fact that
these alterations of the pn-QRPA improved reliability of the 
evaluated matrix elements, further progress of the
nuclear approaches dealing with double beta decay is needed. 

The deformation degrees of freedom of nuclei undergoing the
$2\nu\beta\beta$-decay were first considered within Nilsson
model with pairing \cite{zam82}. Bogdan, Faessler, Petrovici
and Holan calculated the $0\nu\beta\beta$-decay in the pairing 
model including deformation and rotation \cite{bogd85}. 
The first QRPA
calculation of the $2\nu\beta\beta$-decay matrix elements in
a deformed Nilsson-BCS basis were presented in Ref. 
\cite{grotz85}. The authors did not take into account
particle-particle interaction of the nuclear Hamiltonian
and assumed initial and final nuclei to be equally deformed.   
The effects of nuclear deformation were considered also
within the $SU(3)$ scheme \cite{hirsch1}, which has been found 
successfull in describing the heavy  rotational nuclei.
The $SU(3)$ scheme is a tractable shell model theory for
deformed nuclei, which requires a severe truncation of the
single particle basis. This approach was used for calculation
of the double beta decay half-lifes of different heavy 
nuclear systems \cite{hirsch2} 
and the predictions were found to be in good agreement 
with available experimental data for $^{150}Nd$ and $^{238}U$
\cite{data}. The effect of the deformation of the nuclear shape on the 
two-neutrino double beta decay matrix element has been discussed 
in details within a method developed by Raduta, Faessler and Delion 
\cite{rafade}. 
The authors used angular momentum projected single particle basis 
having the energies close to those of Nilsson levels. 
The Gamow-Teller states were generated with help of the spherical 
proton-neutron QRPA with a good angular momentum quantum number
within the considered basis. The results were 
presented for the $2\nu\beta\beta$-decay of $^{82}Se$. It was 
shown that the deformation affects significantly the 
$2\nu\beta\beta$-decay matrix element. The deformation effect on the
double Gamow-Teller matrix element of $^{100}Mo$ were investigated
in the Hartree--Fock--Bogoliubov (HFB) framework in Ref. \cite{dixit}.
It was noticed that there is a necessity of an appropriate amount
of deformation in the HFB intrinsic state to reproduce the
 experimental $2\nu\beta\beta$-decay half-life.  

There is an interest to study the effect of deformation on the
double beta decay matrix elements within the deformed QRPA
\cite{krum,sar98}, which allows unified description of the
$2\nu\beta\beta$-decay in spherical and deformed nuclei. 
This nuclear structure aproach has been found succeessful 
in description of the single $\beta$-decay transitions of 
medium and heavy nuclei a long time ago. In the first
applications only the particle-hole terms of the Gamow-Teller
force were taken into account \cite{krum,sar98}.
It was supposed that the particle-particle terms have minor 
effect on the Gamow-Teller strength function. However, from 
the spherical QRPA calculations we know that the particle-particle 
force plays an important role for describing the $\beta^+$- and 
$\beta\beta$-processes \cite{vogel2,civ1,mut}.
Recently, the importance of the particle-particle interaction
has been confirmed also in the deformed QRPA treatment of the
Gamow-Teller strength distributions \cite{sar01}. A strong 
sensitivity of the single $\beta$-decay characteristics to the 
nuclear shape, RPA ground state correlations and pairing
correlations were manifested. 

The aim of the present paper is to investigate the effect of nuclear
deformation on the double Gamow-Teller matrix element of 
the $2\nu\beta\beta$-decay within the deformed QRPA.
We present here as an example 
numerical results for the $2\nu\beta\beta$-decay of $^{76}Ge$, which is
a prominent transition due to ongoing and planned 
$\beta\beta$ experiments for this isotope \cite{ge76a,zdefu}. 
A subject of our
interest is also the $\beta^-$ and $\beta^+$ distributions
in $^{76}Ge$ and $^{76}Se$, respectively. The problem of 
fixing of the nuclear structure parameters will be addressed.

The paper is organized as follows.
The basic elements of the deformed QRPA formalism are reviewed in
Section 2. In Section 3 we establish our choice for
deformation, pairing and force parameters. Then the results for
the $2\nu\beta\beta$-decay ground state transition 
$^{76}Ge \rightarrow ^{76}Se$ are analyzed in context of the
Gamow-Teller strength distributions of $^{76}Ge$ and 
$^{76}Se$. The final conclusions and remarks are pointed out in Section IV. 

\section{Formalism of the deformed QRPA}
\label{sec: level2}

The theoretical approach is based on the deformed proton-neutron 
Quasiparticle Random Phase Approximation with separable 
proton-neutron residual interaction, which is
relevant for the allowed Gamow-Teller transitions
\cite{krum,sar98,sar01}. 
This approach allows for an unified description of the
$2\nu\beta\beta$-decay process in deformed and spherical nuclei. 

The total nuclear  Hamiltonian takes the form
\begin{equation}
H = H_0 + H_{int}.
\label{eq:1}
\end{equation}
$H_0$ denotes the Hamiltonian for the quasiparticle mean field described by a 
deformed axially-symmetric Woods-Saxon potential \cite{Tan79}
\begin{equation}
H_0 = \sum_{\tau\rho} E_{\tau} a^\dagger_{\tau\rho_\tau} 
a_{\tau\rho_\tau}
(\tau = p,n),
\label{eq:2}
\end{equation}
where $E_{\tau}$ are the quasiparticle energies. $a^+_{\tau\rho_\tau}$
($a^+_{\tau\rho_\tau}$) is the quasiparticle creation (annihilation)
operator. The p (n) index denotes proton (neutron) quasiparticle states 
with projection $\Omega_p$ ($\Omega_n$) of the full angular momentum 
on the nuclear symmetry axis and parity $\pi_p$ ($\pi_n$). The index 
$\rho$ ($\rho = \pm 1$) represents the sign of the angular momentum projection 
$\Omega$. We note that intrinsic states are twofold degenerate. The
states with $\Omega_\tau$ and $-\Omega_\tau$ have the same energy as
consequence of the time reversal invariance. We shall use notation
such that $\Omega_\tau$ is taken to be positive for states and 
negative for time reversed states. 
 
The method includes pairing between like nucleons in the BCS approximation
with fixed gap parameters for protons, $\Delta_p$, and neutrons, $\Delta_n$.
\begin{equation}
~~~E_\tau = \sqrt{(\varepsilon_{\tau} - \lambda_\tau)^2
+ \Delta^2_\tau },
\label{eq:3}
\end{equation}
where $\varepsilon_{\tau}$ are single--particle energies, 
$\Delta_\tau$ is the gap, $\lambda_\tau$ is the Fermi energy.
The Bogolyubov transformation, which defines the 
quasiparticle representation,  is given by
\begin{eqnarray}
\left(\matrix{ c^\dagger_{\tau \rho_\tau} \cr
\tilde{c}_{\tau \rho_\tau}} \right) =
\left( \matrix{ u_\tau & v_\tau \cr
 -v_\tau & u_\tau } \right) 
\left(\matrix{ a^\dagger_{\tau \rho_\tau} \cr
\tilde{a}_{\tau \rho_\tau}} \right). 
\label{eq:4}
\end{eqnarray}
Here, $c^\dagger_{\tau \rho_\tau}$  and
$c_{\tau \rho_\tau}$  are creation and annihilation 
operators for particles, respectively. $\sim$ indicates
the single nucleon state obtained by the time reversal operator 
$T = K e^{-i\sigma_y \pi/2} = \left(\matrix{0 & -1 \cr 1 & 0}\right) K$.
This state is given in the Appendix A \cite{No324}.
The occupation probabilities $v^2_\tau$ are written as
\begin{eqnarray}
v^2_\tau = \frac{1}{2}\left( 1 - 
\frac{\varepsilon_{\tau} - \lambda_\tau}{E_\tau}\right),
~~~ u^2_\tau = 1 - v^2_\tau
\label{eq:5}
\end{eqnarray}
and the proton and neutron number equations are
\begin{equation}
Z = 2 \sum_p v^2_p,  ~~~~~~ N = 2 \sum_n v^2_n,
\label{eq:6}
\end{equation}
where Z and N are the numbers of protons and neutrons, 
respectively.

The residual interaction part $H_{int}$ of
nuclear Hamiltonian in Eq. (\ref{eq:1})
contains two terms associated with 
particle--hole (ph) and particle--particle
(pp) interaction:
\begin{eqnarray}
H_{int} =
~~\chi \sum_{K=0,\pm 1}
  (-1)^K ( \beta^-_{1K}\beta^+_{1-K}
+ \beta^+_{1-K} \beta^-_{1K}) ~
- \kappa \sum_{K=0,\pm 1} 
  (-1)^K ( P^-_{1K}P^+_{1-K}
+ P^+_{1-K} P^-_{1K} ).
\label{eq:7}
\end{eqnarray}
The operators $\beta^-$ and $P^-$ are ph and pp components 
of the spin-isospin $\tau^+\sigma$, namely,
\begin{eqnarray}
\beta^-_K &=& \sum_{p \rho_p n \rho_n}
<p \rho_p| \tau^+ \sigma_K |n \rho_n> 
c^\dagger_{p \rho_p} c_{n \rho_n}, ~~~\beta^+_K = (\beta^-_K)^\dagger, 
\nonumber \\
P^-_K &=& \sum_{p \rho_p n \rho_n}
<p \rho_p| \tau^+ \sigma_K |n \rho_n> 
c^\dagger_{p \rho_p} {\tilde c}^\dagger_{n \rho_n}, 
~~~P^+_K = (P^-_K)^\dagger. 
\label{eq:8}
\end{eqnarray}
The ph and pp forces in Eq. (\ref{eq:6}) are defined to be 
repulsive and attractive ($\chi, \kappa \ge 0$), respectively,
reflecting the general feature of the  nucleon-nucleon interaction 
in the $J^\pi=1^+$ channel. The explicit form of the matrix element
$<p \rho_p| \tau^+ \sigma_K |n \rho_n> $ is presented in the Appendix A. 

After neglecting the scattering terms 
$a^\dagger_{p \rho_p}a_{n \rho_n}$ and
$a^\dagger_{n \rho_n}a_{p \rho_p}$ the
quasiparticle representation of $H_{int}$ takes the form
\begin{eqnarray}
H_{int} &=& ~~~ \chi \sum_{K=0,\pm 1} \sum_{i j}
[ ( \sigma_K(i) {\overline{\sigma}}_K(j)+
 {\overline{\sigma}}_K(i)  \sigma_K(j))
(A^\dagger (i, K) A^\dagger ({\bar j}, K) +  
A ({\bar j},K) A(i,K) )
\nonumber \\
&&~~~~~~~~~~+~ ( \sigma_K(i) {{\sigma}}_K(j)+
 {\overline{\sigma}}_K(i)  {\overline{\sigma}}_K(j))
(A^\dagger (i, K) A(j, K) +  A(j, K) A^\dagger (i,K) )]
\nonumber \\
&& ~ - \kappa \sum_{K=0, \pm 1} \sum_{i j}
[ - ( \pi_K(i) {\overline{\pi}}_K(j)+
 {\overline{\pi}}_K(i)  \pi_K(j))
(A^\dagger (i, K) A^\dagger ({\bar j}, K) +  
A({\bar j}, K) A(i, K) )
\nonumber \\
&&~~~~~~~~~~+~ ( \pi_K(i) {{\pi}}_K(j)+
 {\overline{\pi}}_K(i)  {\overline{\pi}}_K(j))
(A^\dagger (i, K) A(j, K) +  A(j, K) A^\dagger (i, K) )],
\label{eq:9}
\end{eqnarray}
with 
\begin{eqnarray}
\sigma_K(i) &=& <p \rho_p|\tau^+ \sigma_K|n \rho_n > u_p v_n,
~~{\overline{\sigma}}_K(i) = 
<p \rho_p|\tau^+ \sigma_K|n \rho_n > v_p u_n, \nonumber \\
\pi_K(i) &=& <p \rho_p|\tau^+ \sigma_K|n \rho_n > u_p u_n,
~~{\overline{\pi}}_K(i) = 
<p \rho_p|\tau^+ \sigma_K|n \rho_n > v_p v_n. 
\label{eq:10}
\end{eqnarray}
$A^\dagger_i$ and $A_i$  are the two-quasiparticle creation
and annihilation operators
\begin{equation}
A^\dagger (i, K) = a^\dagger_{p \rho_p} a^\dagger_{n \rho_n}, ~~
A^\dagger ({\bar i}, K) = 
{\tilde a}^\dagger_{p \rho_p} {\tilde a}^\dagger_{n \rho_n}, 
~~A(i, K) = (A^\dagger (i, K))^\dagger.
\label{eq:11}
\end{equation}
The quasiparticle pairs $i$ and ${\bar i}$ are defined by the selection
rules $\Omega_p - \Omega_n = K$ and $\Omega_n - \Omega_p = K$, 
respectively, and $\pi_p \pi_n = 1.$ 

The above considered model Hamiltonian includes terms with $K=0,\pm 1$
and describes $J^\pi K = 1^+ 1, 1^+0$ excitations. In the laboratory
frame the proton-neutron QRPA phonon   wave functions for
Gamow-Teller excitations in even-even nuclei have the form 
\begin{eqnarray} 
|1M(K),m > &=& \sqrt{\frac{3}{16 \pi^2}}
[ {\cal D}^1_{M K}(\phi,\theta,\psi) {Q^m_K}^\dagger + 
(-1)^{1+K} {\cal D}^1_{M -K}(\phi,\theta,\psi) 
{Q^m_{-K}}^\dagger ] |rpa> ~~~(K=\pm1),
\nonumber \\
|1M(K),m > &=& \sqrt{\frac{3}{8 \pi^2}}
{\cal D}^1_{M K}(\phi,\theta,\psi) {Q^m_K}^\dagger |rpa>
~~~(K=0),
\label{eq:12}
\end{eqnarray}
where $|rpa>$ denotes the QRPA ground state. The intrinsic states are
generated by the phonon creation operator
\begin{equation}
{Q^m_K}^\dagger = 
\sum_i [ X^{m}_{i, K} A^\dagger (i, K) - 
Y^{m}_{i, K} A({\bar i}, K) ].
\label{eq:13}
\end{equation}
In the case $K=\pm 1$ ($K=0$) the sum in Eq. (\ref{eq:13}) 
includes  all bound and quasibound two-quasiparticle 
spin--projection--flip (non--spin--projection--flip) 
configurations. We note that the 
$K = -1$ and $K = 1$ modes are related to each other 
through time reversal and are degenerate. 

The excitation energy $\omega_k$ and the amplitudes 
$X^m_{i, K}$ and $Y^m_{i, K}$ of the phonon ${Q^m_K}^\dagger$
are obtained by solving the RPA matrix equation
\begin{eqnarray}
\left( \matrix{ {\cal A}(K) & {\cal B}(K) \cr
{\cal B}(K) & {\cal A}(K) }\right) 
~\left( \matrix{ X^m_K \cr Y^m_K} \right)~ = ~
\omega^m_K ~
\left( \matrix{ 1 & 0 \cr 0 & -1 }\right) 
~\left( \matrix{ X^m_K \cr Y^m_K} \right),
\label{eq:14}
\end{eqnarray}
where 
\begin{eqnarray}
{\cal A}_{ij}(K) &=& {\cal E}_i \delta_{i j}
+ 2 \chi [ \sigma_K (i) \sigma_K (j) + 
{\overline{\sigma}}_K(i) {\overline{\sigma}}_K(j)] 
- 2 \kappa [ \pi_K (i) \pi_K (j) + 
{\overline{\pi}}_K(i) {\overline{\pi}}_K(j)], \nonumber \\
{\cal B}_{ij}(K) &=& ~~~~~~~~~~~
 2 \chi [ \sigma_K (i) {\overline{\sigma}}_K(j) +
{\overline{\sigma}}_K(i)  \sigma_K (j) ]
+ 2 \kappa [ \pi_K (i) {\overline{\pi}}_K(j) +
{\overline{\pi}}_K(i)  \pi_K (j) ]
\label{eq:15}
\end{eqnarray}
with  ${\cal E}_i = E_p + E_n$ the two-quasiparticle
excitation energy.

An advantage of using the separable forces  is that the 
RPA matrix equation reduces to a homogeneous system of
only four equations for the four unknown norms
$N_\sigma$, $N_{\bar{\sigma}}$, $N_\pi$ and $N_{\bar{\pi}}$,
which is much easier to solve in comparison 
with the full diagonalization of RPA matrix of large 
dimension. The corresponding secular equation is given 
by
\begin{equation}
\left| \matrix{ 
1+\chi (-P^K_{\sigma\sigma}+R^K_{\bar{\sigma}\bar{\sigma}} ) ~&~
\chi ( -P^K_{\sigma\bar{\sigma}}+R^K_{\bar{\sigma}\sigma} ) ~&~
\kappa (P^K_{\sigma\pi}+R^K_{\bar{\sigma}\bar{\pi}} ) ~&~
\kappa (P^K_{\sigma\bar{\pi}}+R^K_{\bar{\sigma}\pi} ) \cr
\chi ( -P^K_{\bar{\sigma}\sigma}+R^K_{\sigma\bar{\sigma}} ) ~&~
1+\chi (-P^K_{\bar{\sigma}\bar{\sigma}}+R^K_{\sigma\sigma} ) ~&~
\kappa (P^K_{\bar{\sigma}\pi}+R^K_{\sigma\bar{\pi}} ) ~&~
\kappa (P^K_{\bar{\sigma}\bar{\pi}}+R^K_{\sigma\pi} ) \cr
\chi (-P^K_{\pi\sigma}-R^K_{\bar{\pi}\bar{\sigma}} ) ~&~
\chi ( -P^K_{\pi\bar{\sigma}}-R^K_{\bar{\pi}\sigma} ) ~&~
1+\kappa (P^K_{\pi\pi}-R^K_{\bar{\pi}\bar{\pi}} ) ~&~
\kappa (P^K_{\pi\bar{\pi}}-R^K_{\bar{\pi}\pi} ) \cr
\chi ( -P^K_{\bar{\pi}\sigma}-R^K_{\pi\bar{\sigma}} ) ~&~
\chi (-P^K_{\bar{\pi}\bar{\sigma}}-R^K_{\pi\sigma} ) ~&~
\kappa (P^K_{\bar{\pi}\pi}-R^K_{\pi\bar{\pi}} ) ~&~
1+\kappa (P^K_{\bar{\pi}\bar{\pi}}-R^K_{\pi\pi} ) 
 } \right| = 0, 
\label{eq:16}
\end{equation}
with
\begin{equation}
P^K_{\alpha\acute{\alpha}} = 
2 \sum_i \frac{\alpha (i) \acute{\alpha}(i)}{\omega^m_K - {\cal E}_i},
~~~~~
R^K_{\alpha\acute{\alpha}} = 
2 \sum_i \frac{\alpha (i) \acute{\alpha} (i)}{\omega^m_K + {\cal E}_i},
~~~~(\alpha,\acute{\alpha} = \sigma, \bar{\sigma},
\pi, \bar{\pi}).
\label{eq:17}
\end{equation}
The forward and backward amplitudes are written as
\begin{eqnarray}
X^m_{i K} &=& \frac{2 N_\sigma}{\omega^m_K - {\cal E}_i}[~
\chi ( {\sigma_K (i)}  +
 {{\bar{\sigma}}_K (i)} \frac{N_{\bar{\sigma}}}{N_\sigma} ) -
\kappa ( {\pi_K (i)} \frac{N_\pi}{N_\sigma} + 
 {{\bar{\pi}}_K (i)} \frac{N_{\bar{\sigma}}}{N_\sigma} )~],
\nonumber \\ 
Y^m_{i K} &=& \frac{-2 N_\sigma}{\omega^m_K + {\cal E}_i}[~
\chi ( {{\bar{\sigma}}_K (i)}  +
 {{{\sigma}}_K (i)} \frac{N_{\bar{\sigma}}}{N_\sigma} ) +
\kappa ( {{\bar{\pi}}_K (i)} \frac{N_\pi}{N_\sigma} + 
 {{{\pi}}_K (i)} \frac{N_{\bar{\sigma}}}{N_\sigma}  )~],
\label{eq:18}
\end{eqnarray}
where for the norms $N_\sigma$, $N_{\bar{\sigma}}$, $N_\pi$ and $N_{\bar{\pi}}$
we have
\begin{eqnarray}
N_\sigma &=& \sum_j [\sigma_K(i) X^m_{j K} + 
{\bar{\sigma}}_K(i) Y^m_{j K}],
~~~
N_{\bar{\sigma}} = \sum_j [{\bar{\sigma}}_K(i) X^m_{j K} + 
\sigma_K(i) Y^m_{j K}],
\nonumber \\
N_\pi &=& \sum_j [\pi_K(i) X^m_{j K} - 
{\bar{\pi}}_K(i) Y^m_{j K}],
~~~
N_{\bar{\pi}} = \sum_j [{\bar{\pi}}_K(i) X^m_{j K} - 
\pi_K(i) Y^m_{j K}].
\label{eq:19}
\end{eqnarray}
The normalization factor $N_{\sigma}$ is determined from the
condition
\begin{equation}
<rpa| [Q^m_K, {Q^m_K}^\dagger ] |rpa> =
\sum_i ( X^m_{i K} X^m_{i K} - Y^m_{i K} Y^m_{i K}) = 1.
\label{eq:20}
\end{equation} 
We note that the QRPA equations are calculated separately
for different values of K and that the solutions for
$K=+1$ and $K=-1$ coincide to each other due to 
considered axial symmetry. 
 
The $\beta^-$ and $\beta^+$ transition amplitudes from $0^+$
even-even initial nuclear state to a one-phonon
state in odd-odd final nucleus are expressed by
\begin{eqnarray}
<1M(K),m| {\hat{\beta}}^-_M |0^+_{g.s.}> &=& 
\sum_i [ \sigma(i) X^m_{i K} + {\bar{\sigma}}(i) Y^m_{i K} ],
\nonumber \\
<1M(K),m| {\hat{\beta}}^+_M |0^+_{g.s.}> &=& 
\sum_i [ \bar{\sigma}(i) X^m_{i K} + \sigma (i) Y^m_{i K} ].
\label{eq:21}
\end{eqnarray}
Here, $|0^+_{g.s.}>$ denotes the 
correlated RPA ground state in the laboratory frame. 
The ${\hat{\beta}}^\pm$
transition operators in Eq. (\ref{eq:21}) are related with 
intrinsic $\beta^{\pm}$ operators in Eq. (\ref{eq:7}) as follows:
\begin{equation}
{\hat{\beta}}^\pm_{M} = \sum_\mu {\cal D}^1_{M \mu}
(\phi,\theta,\psi) \beta^\pm_\mu.
\label{eq:22}
\end{equation}
For the $\beta$ strength function we have
\begin{eqnarray}
B^-_{GT}(\omega ) &=& \sum_{K=0,\pm 1} \sum_{ m}
|<1(K),m\parallel {\hat{\beta}}^- \parallel0^+_{g.s.}>|^2 
\delta(\omega - \omega_K), \nonumber \\
B^+_{GT}(\omega ) &=& \sum_{K=0,\pm 1} \sum_{ m}
|<1(K),m\parallel {\hat{\beta}}^+ \parallel0^+_{g.s.}>|^2 
\delta(\omega - \omega_K).
\label{eq:23}
\end{eqnarray}

From the $\beta^\pm$ amplitudes one obtains, straightforwardly,
the total $\beta^\pm$ strengths:
\begin{eqnarray}
S^-_{GT} &=& \sum_{K=0,\pm 1} \sum_{ m}
|<1(K),m\parallel {\hat{\beta}}^- \parallel0^+_{g.s.}>|^2, ~~~ 
\nonumber \\
S^+_{GT} &=& \sum_{K=0,\pm 1} \sum_{ m}
|<1(K),m\parallel {\hat{\beta}}^+ \parallel0^+_{g.s.}>|^2 
\label{eq:24}
\end{eqnarray}
and for the Ikeda sum rule we get
\begin{eqnarray}
S^-_{GT} - S^+_{GT} &=& 
\sum_{K=0,\pm 1} [ (\sigma_K (i))^2 - ({\bar{\sigma}}_K (i))^2 ]  
\nonumber \\
&=& \sum_{K=0,\pm 1} \sum_{p \rho_p n \rho_n}
|<p \rho_p| \tau^+ \sigma_K | n \rho_n>|^2 (v_n^2 - v_p^2) = 3 (N-Z).
\label{eq:25}
\end{eqnarray}
The factor 3 comes from the sum over $K$, i.e., the contribution
from each component K is equal to (N-Z). In deriving the above
expression we used the closure condition for QRPA states, 
assumed the completeness relation for single particle states
\begin{equation}
\sum_{\tau \rho_\tau} |\tau \rho_\tau><\tau \rho_\tau|=1 
\label{eq:26}
\end{equation}
and applied Eq. (\ref{eq:6}). We note that if a truncated single particle 
basis is considered in respect to that the full single particle basis of the
Woods-Saxon potential, the condition in Eq. (\ref{eq:26})
is violated and as a consequence the Ikeda sum rule as well.

The inverse half-life of the $2\nu\beta\beta$-decay  can
be expressed as a product of  an accurately known 
 phase-space factor $G^{2\nu}$ 
and the Gamow-Teller transition matrix element $M^{2\nu}_{GT}$ 
in second order:
\begin{equation}
[ T^{2\nu}_{1/2}(0^+_{g.s.} \rightarrow 0^+_{g.s.}) ]^{-1} = 
G^{2\nu} ~(g_A)^4~ | M^{2\nu}_{GT}|^2.
\label{eq:27}
\end{equation}
The contribution from the two successive Fermi transitions is
safely neglected as they come from isospin mixing effect \cite{hax}.
Within the deformed QRPA approach 
the double Gamow-Teller matrix element $M^{2\nu}_{GT}$ 
for ground state to ground state $2\nu\beta\beta$-decay 
transition takes the form
\begin{equation}
M^{2\nu}_{GT}=\sum_{{m_i m_f}} \sum_{K=0,\pm 1}
\frac{<0^+_f\parallel \beta^- \parallel 1(K),m_f>
<1(K),m_f|1(K),m_i>
<1(K),m_i\parallel \beta^- \parallel 0^+_i>}
{(\omega^{m_f}_K + \omega^{m_i}_K)/2}.
\label{eq:28}
\end{equation}
The sum extends over all $1^+$ states of the intermediate nucleus.
The index i (f) indicates that the quasiparticles and the excited
states of the nucleus are defined with respect to the initial (final)
nuclear ground state $|0^+_i>$ ($|0^+_f>$). The overlap is necessary since
these intermediate states are not orthogonal to each other.
The two sets of intermediate nuclear states generated from the 
initial and final ground states are not identical within the
considered approximation scheme. 
Therefore the overlap factor of these states $<1(K),m_f|1(K),m_i>$
is introduced in the theory \cite{pan88}. Its takes the form
\begin{equation}
<1(K),m_f|1(K),m_i> =
\sum_{l_i l_f}
~[X^{m_f}_{l_f K}X^{m_i}_{l_i K}-Y^{m_f}_{l_f K}Y^{m_i}_{l_i K}] 
~{\cal R}_{l_f l_i} 
~<BCS_f|BCS_i>. 
\label{eq:29}
\end{equation}
The factor ${\cal R}_{l_f l_i}$, which includes the overlaps of single particle 
wave functions of the initial and final nuclei is given explicitely in the 
Appendix B. There, a detailed derivation of ${<1(K),m_f|1(K),m_i>}$
together with the overlap factor of the initial and final 
BCS vacua can be find as well. In the spherical limit the value of the BCS 
overlap factor is about 0.8 and it is worth mentioning that it
was comonnly neglected in the double beta 
decay calculations \cite{civ1,mut,suho00,grif92}.  

We note that we neglected the overlap matrix elements 
between the intermediate
states generated from initial and final nuclei with different K
as the corresponding factors are very small due to different
structure of corresponding RPA configurations. Thus the 
spin--projection--flip
($K=\pm 1$) and non--spin--projection--flip ($K=0$) excitations contribute
coherently to the $2\nu\beta\beta$-decay matrix element 
$M^{2\nu}_{GT}$.

\section{Calculation and Discussion}

The formalism described in Section II  is used for the calculation
of the $2\nu\beta\beta$-decay ground state transition
$^{76}Ge \rightarrow {^{76}Se}$. The results are obtained
with a deformed, axially symmetric  Woods-Saxon potential
\cite{Da135}. The deformation independent Woods-Saxon parameters
(well depth, skin thickness, radius and spin--orbit constants)
are taken from \cite{Tan79}. This parameterization of the
Woods-Saxon potential was used previously
in different RPA calculations, where a good agreement was achieved
with experimental data, in particular  for single M1 transitions
at low energy, observed in $(e,e')$ and $(\gamma,\gamma')$ experiments
\cite{Noj93,Sar93}.

The quadrupole-hexadecapole parameterization of the nuclear surface 
is considered:
\begin{equation}
R(\theta) = R_n [1 + \beta_2 Y_{20}(\theta) + \beta_4 Y_{40} (\theta)],
\label{eq:31}
\end{equation} 
where $R_n$ is the normalizing constant ($R(\theta=0)=1$) and $Y_{L0}$
($L=2,4$) are spherical harmonics. The spherical limit is achieved 
for $\beta_2=\beta_4=0$.

The deformation parameter $\beta_2$ can be deduced from nuclear 
electric quadrupole laboratory moment ($Q_p = -(7/2) Q$, $Q$ and $Q_p$ are
laboratory and intrinsic quadrupole moments, respectively)
or extracted from values based on measured E2 probability 
($Q_p = \sqrt{16 \pi B(E2) /5e^2}$, the sign can not be extracted)
via the intrinsic quadrupole moments $Q_p$: 
$\beta_2 = \sqrt{\pi/5} Q_p/(Z r_c^2)$ ($r_c$ is the charge root
mean square radius). However, the available experimental
data associated with the above two approaches \cite{balraj,ragha,raman} 
lead to a different quadrupole deformation both for  $^{76}Ge$
and $^{76}Se$. In particular, the quadrupole moments measured by 
Coulomb excitation reorientation method \cite{ragha},
which yields also the sign of the intrinsic quadrupole moment,
imply $\beta_2 = 0.1$ ($^{76}Ge$) and $\beta_2 = 0.16$ ($^{76}Se$) 
\cite{pedro},
but from  the measured  values of $B(E2)$ strengths \cite{raman}
one finds $|\beta_2| = 0.26$ ($^{76}Ge$) and  $|\beta_2| = 0.29 $ 
($^{76}Se$) \cite{pedro}. A lack of accurate experimental information
on the deformation of $^{76}Ge$  and $^{76}Se$ suggests that 
it is necessary to study the associated $2\nu\beta\beta$-decay matrix element 
as a function of deformation parameters of both initial and final
nuclei. 

We note that there are theoretical calculations
of deformation parameters performed within relativistic mean field theory 
\cite{ring}, which imply
\begin{equation}
^{76}Ge:~\beta_2 = 0.157,~ \beta_4 = ~ 0.019, ~~
^{76}Se:~\beta_2 = -0.244,~\beta_4 = -0.028.
\label{eq:32}
\end{equation}
These results are in good agreement with  predictions of the 
macroscopic-microscopic model of M\"oller, Nix, Myers and Swiatecki 
\cite{moll} as well as with nonrelativistic 
Hartree--Fock calculations \cite{pedro}.   

In the deformed QRPA calculation we assume a truncated model space by
considering only single particle states with  
maximal allowed value of the asymptotic  quantum
number $N$ equal to 5. In contrast to many other microscopic 
calculations, energies of single particle levels are not shifted but 
taken exactly as provided by the deformed Woods-Saxon potential. 

\begin{figure}
\centerline{\psfig{figure=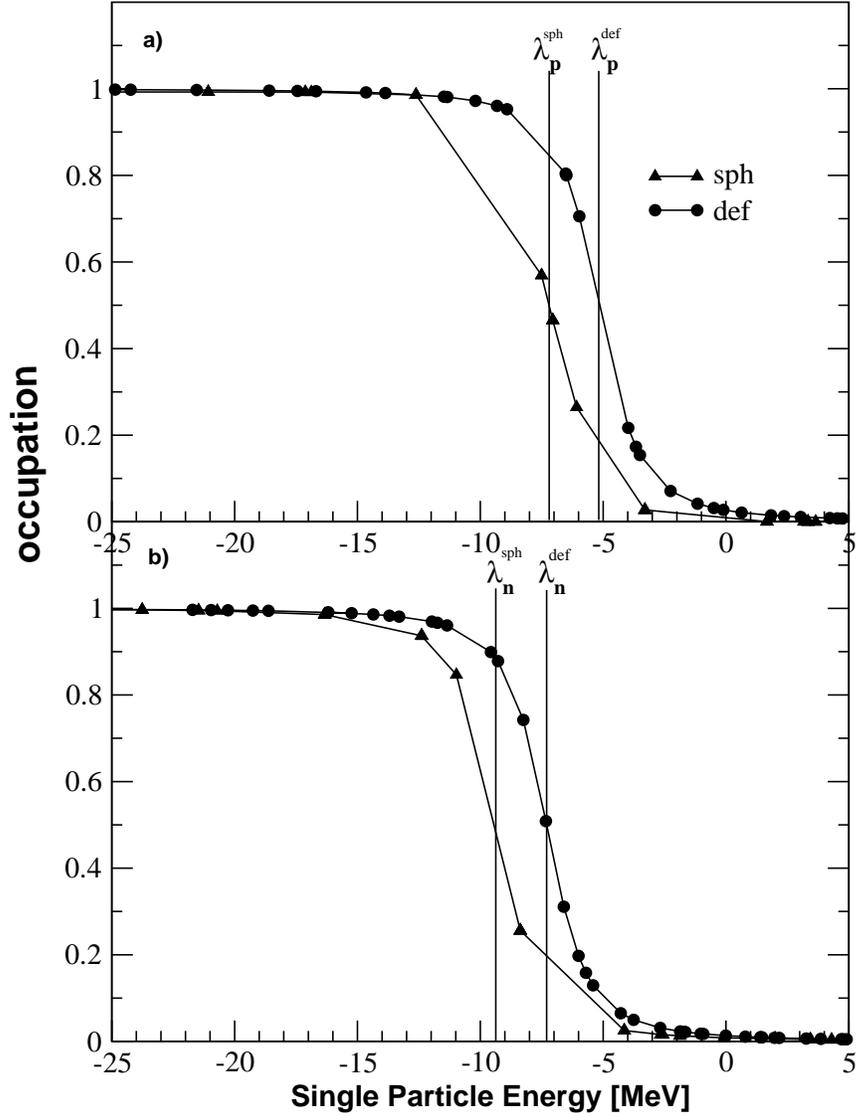,width=14cm,height=18cm}}
\caption{
The proton (a) and neutron (b) occupation probabilities 
close to the Fermi level for $^{76}Se$. The spherical 
(sph, triangle up) and deformed (def, closed circle) 
BCS results correspond to pairing gaps in Eq. (\ref{eq:33}).
The vertical lines denote Fermi energy for protons
($\lambda_p$) and neutrons ($\lambda_n$).}
\label{fig.1}
\end{figure}

The BCS equations are solved for protons and neutrons. 
The proton and neutron pairing gaps are determined 
phenomenologically to reproduce the odd-even mass
differences through a symmetric five-term formula \cite{waps}.
For both spherical and deformed shapes of the studied 
A=76 nuclei we consider the same values of the
 gap parameters, which are given by 
\begin{equation}
^{76}Ge:~\Delta_p = 1.561~MeV,~\Delta_n = 1.535~MeV, ~
^{76}Se:~\Delta_p = 1.751~MeV,~\Delta_n = 1.710~MeV.
\label{eq:33}
\end{equation}
In Fig. \ref{fig.1} the calculated proton and neutron occupation 
probabilities close to the Fermi level are presented for $^{76}Se$.
We see that both for protons and neutrons the BCS solutions 
associated with spherical and deformed nuclear shapes 
differ significantly each from other. This deformation dependence
of the BCS results is expected to have strong impact on the
QRPA solution.

The calculation of the QRPA energies and wave functions requires  
the knowledge of the particle-hole $\chi$ and particle--particle
$\kappa$ strengths of the residual interaction. The optimal 
value of $\chi$ is determined by reproducing the  systematics
of the empirical position of the Gamow-Teller giant resonance
in odd-odd intermediate nucleus as obtained from the $(p,n)$
reactions \cite{homa,taig}. 
The parameter $\kappa$ can be determined  by exploiting 
the systematics of single $\beta$-decay feeding the initial and final
nuclei. In Ref. \cite{homa} the strengths of the particle--hole and 
particle--particle terms of the separable Gamow-Teller force 
were fixed as smooth functions of mass number A by reproducing 
the $\beta$-decay properties of nuclei up to A=150 
within the spherical pn-QRPA model.
The  recommended values of $\chi$ and $\kappa$ of Ref. \cite{homa} 
are as follows:
\begin{equation}
\chi = 5.2/A^{0.7}~MeV, ~~~~~\kappa = 0.58/A^{0.7}~MeV.
\label{eq:34}
\end{equation}
In the presented  $2\nu\beta\beta$-decay  study of the ground
state transition $^{76}Ge\rightarrow {^{76}Se}$  we  use
the recommended value for $\chi$ ($\chi = 0.25~MeV$)
and $\kappa$ is considered as free variable.

\begin{figure}
\centerline{\psfig{figure=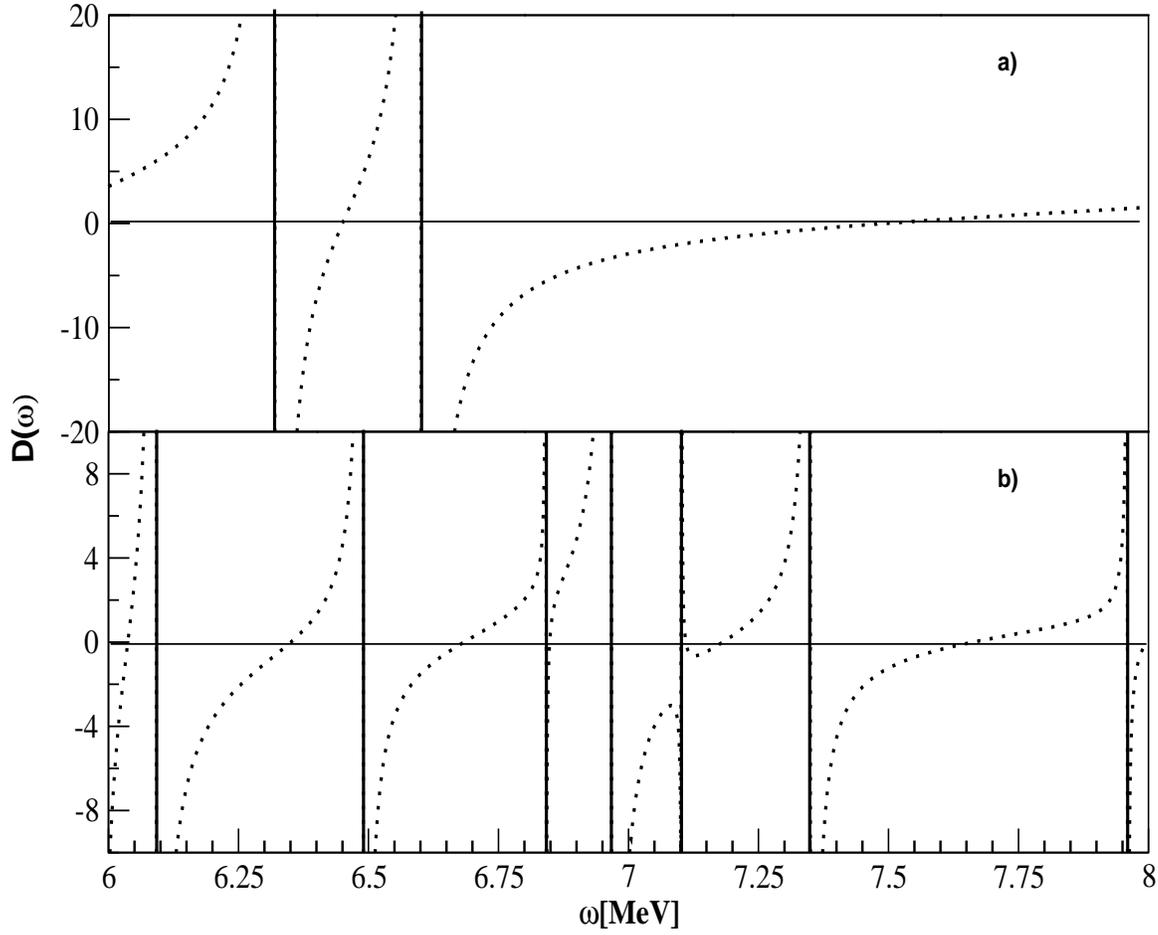,width=16cm,height=18cm,angle=-90}}
\caption{
The l.h.s. of Eq. (\ref{eq:16}) $D(\omega)$
is plotted as function of the energy $\omega$ for $^{76}Ge$.
The calculation has been performed within spherical (a) and
deformed  (b) QRPA for $\chi = 0.25$ and $\kappa = 0$. 
In case (b) the K=0 results are presented.
}
\label{fig.2}
\end{figure}

The QRPA calculations for the $K=0, \pm 1$ states are performed 
by following the procedure described in Section II. 
The number of zeros of the non-linear  RPA secular equation in Eq. 
(\ref{eq:16}) is equal to the number of the two-quasiparticle configurations
included in the microscopic sums (\ref{eq:17}). 
In the case of  $K=0$ ($K=\pm 1$) the dimension of the configuration 
space is of the order of  450 (900) 
quasiparticle pairs excitations. To each zero of the 
RPA secular equation corresponds one RPA energy $\omega^m_K$.
It is worthwhile to notice that by introducing the deformation degrees 
of freedom the zeros are not found in each subinterval of energy $\omega$
determined  by the two subsequent  two-quasiparticle excitation energies 
and that in some subintervals two or more zeros are found.
This situation is illustrated
in Fig. \ref{fig.2}. The vertical axis is the left-hand side (l.f.s.)
of Eq.(\ref{eq:16}) and the horizontal one is the energy $\omega_{K=1}$
in the range $6~MeV \le \omega_{k=1} \le 8~MeV$. The upper (a) and lower (b) 
subfigures show results for A=76 obtained within
the spherical and deformed  QRPA, respectively. 
We note that a similar phenomenon  was  found 
also in Ref. \cite{radu84} in different context.

\begin{figure}
\centerline{\psfig{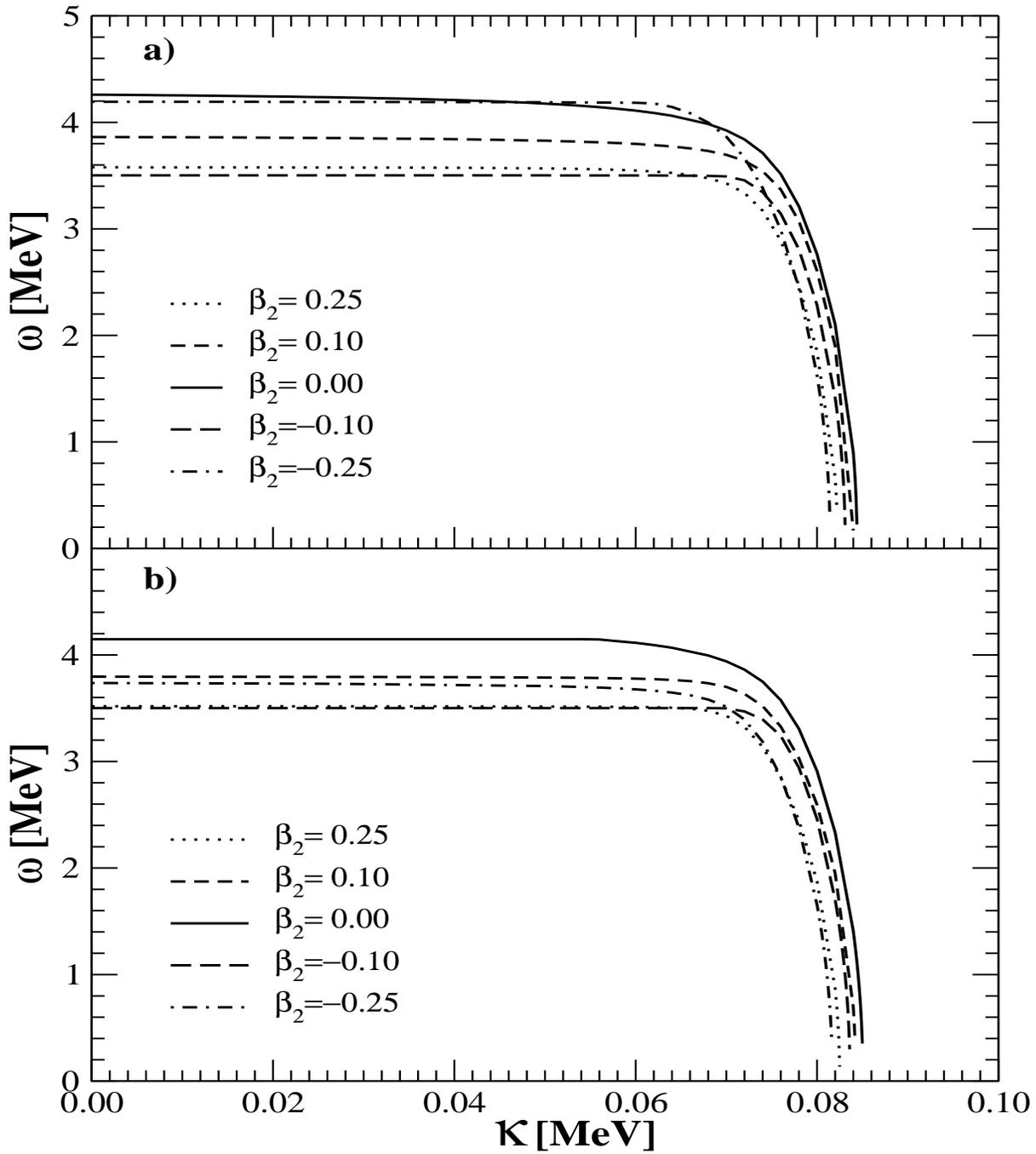}}
\vspace{1cm}
\caption{The energy of the lowest QRPA state as function
of particle--particle interaction strength $\kappa$. 
The K=0 and K=1 deformed QRPA results for $^{76}Se$ are 
presented in subfigures (a) and (b), respectively.
}
\label{fig.3}
\end{figure}

The main drawback of the QRPA is the overestimation 
of the ground state correlations  leading to a collapse of the 
QRPA ground state, which  may lead to ambiguous determination 
of the $\beta$ and $\beta\beta$-decay matrix elements. 
The QRPA collapses, because it generates too many correlations 
with increasing strength of the attractive proton-neutron interaction
as a result of the quasiboson approximation used.  
An interesting issue is the dependence of the position
of the collapse of the QRPA solution on the deformation degrees 
of freedom. In Fig. \ref{fig.3} the energies of the lowest 
$K=0$ and $|K| = 1$ states in $^{76}As$ calculated within the 
deformed QRPA in respect to the $^{76}Se$ ground state
are plotted as  a function of particle-particle interaction strength
$\kappa$. The calculations were performed for different values of
quadrupole deformation parameter $\beta_2$, namely 
$\beta_2 = -0.25, -0.10$, $\beta_2 = 0.0$ and
$\beta_2 = 0.10, 0.25$ representing oblate, spherical and prolate
shapes of $^{76}Se$, respectively. We note that in the spherical limit the
energies of $K = 0$ and $|K| = 1$ states of the intermediate
nucleus coincide to each other. Fig. 
\ref{fig.3} shows that  for a deformed nucleus the collapse of 
the QRPA solution appears for a slightly smaller value of $\kappa$.
In the case of  strongly deformed $^{76}Se$ 
this effect is more pronounced  
in comparison with the less deformed one. 
Thus the deformation degrees of freedom  do not improve 
the stability of the QRPA solution. 
The violation of the Pauli exclusion principle affects 
the QRPA results strongly in the range of particle-particle
interaction $\kappa$ above $0.07$ MeV.

The half-life of the $2\nu\beta\beta$-decay of $^{76}Ge$
is known with high accuracy from the Heidelberg-Moscow 
experiment, in particular 
$T^{2\nu}_{1/2} = [1.55 \pm 0.01 (stat)^{+0.19}_{-0.15}syst.)]
\times 10^{21}$ years \cite{ge76a}. All existing positive results
on the $2\nu\beta\beta$-decay were analyzed  by Barabash  \cite{barab},
who suggested to consider for the ground state transition 
$^{76}Ge \rightarrow {^{76}Se}$ the average value 
$T^{2\nu}_{1/2} = 1.43^{+0.19}_{-0.07}\times 10^{21}$ years. 
By using of Eq. (\ref{eq:27})
and the knowledge of the kinematical factor $G^{2\nu}$
[$G^{2\nu}(^{76}Ge) = 1.49\times 10^{-20}~year^{-1}~MeV^2$]
from the $2\nu\beta\beta$-decay half-life of $^{76}Ge$ one
can deduce the absolute value of the nuclear matrix element 
$|M_{GT}^{2\nu-exp.}|$ equal to $0.138~MeV^{-1}$
by assuming $g_A=1.25$.  
If the value of the axial coupling constant $g_A$
is considered to be unity, the value of $|M_{GT}^{2\nu-exp.}|$ deduced
from the average half-life for A=76 is larger,  $0.216~MeV^{-1}$.

\begin{figure}
\centerline{\psfig{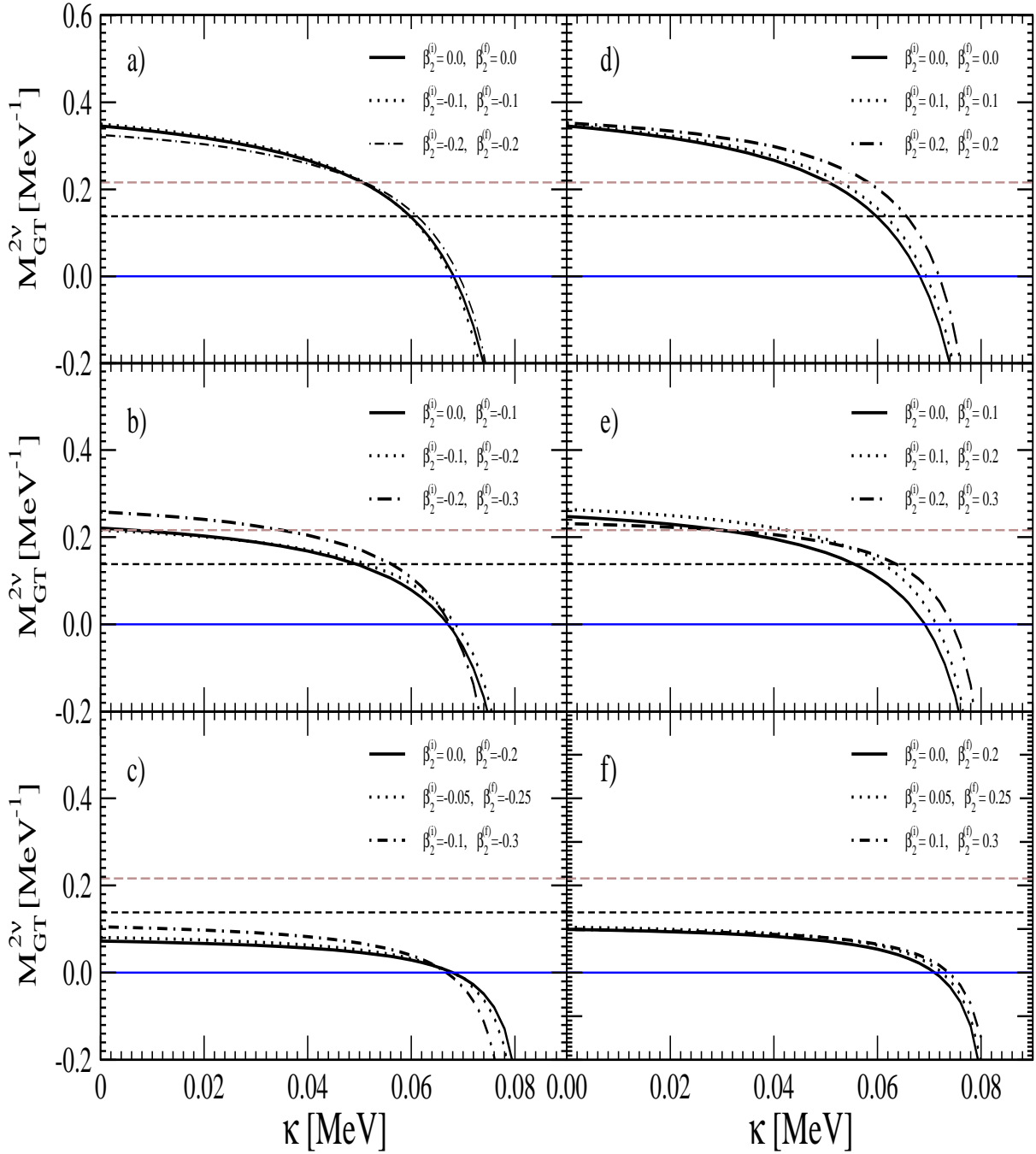}}
\caption{
$2\nu\beta\beta$-decay matrix element of $^{76}Ge$ as function of 
particle--particle interaction strength $\kappa$. 
In the subfigures (a), (b) and (c) [(d), (e) and (f)] the results 
corresponding to oblate [prolate] deformation of both initial and
final nuclei are presented. Please note that if the deformation of the 
initial and final nuclei is comparable, there is only minimal difference
between the calculated values of $M^{2\nu}_{GT}$. 
By increasing difference in deformations of parent and daughter nuclei  
the suppression of $M^{2\nu}_{GT}$ is increased in the range
$0~MeV\le \kappa \le 0.6~MeV$. The two dashed horizontal lines
correspond to $M^{2\nu-exp}_{GT} = 0.216~MeV^{-1}$ ($g_A = 1.0$) and
$M^{2\nu-exp}_{GT} = 0.138~MeV^{-1}$ ($g_A = 1.25$). 
} 
\label{fig.4}
\end{figure}

In Fig. \ref{fig.4} the effect of the deformation on the 
$2\nu\beta\beta$-decay matrix element $M_{GT}^{2\nu}$ is 
analyzed. The results for the $2\nu\beta\beta$-decay 
matrix element are displayed as function of the 
particle--particle strength $\kappa$. The curves drawn 
in subfigures a), b) and c) [subfigures d), e) and f)]  
correspond to the case both initial and final nuclei 
are oblate [prolate].  The two horizontal lines
represent $M_{GT}^{2\nu-exp.}=0.138~MeV^{-1}~(g_A=1.25)$
and $M_{GT}^{2\nu-exp.}=0.216~MeV^{-1}~(g_A=1.0)$. 
By glancing  Fig. \ref{fig.4} 
we note that within the whole range of $\kappa$
there is only a minimal difference between values
of  $M_{GT}^{2\nu}$ corresponding to  the same value of 
$\Delta\beta_2$, which is defined as 
\begin{equation}
\Delta\beta_2 = |\beta_2({^{76}Ge}) - \beta_2({^{76}Se})|.
\end{equation}
In addition, one finds that by increasing the value of $\Delta\beta_2$
the suppression of $M_{GT}^{2\nu}$ becomes stronger within the  range
$0~MeV~\le \kappa \le 0.06~MeV$. This a new suppression mechanism
of the $2\nu\beta\beta$-decay matrix element namely, 
$M_{GT}^{2\nu}$ depends strongly on difference
in deformations of parent and daugter nuclei.  

\begin{figure}
\centerline{\psfig{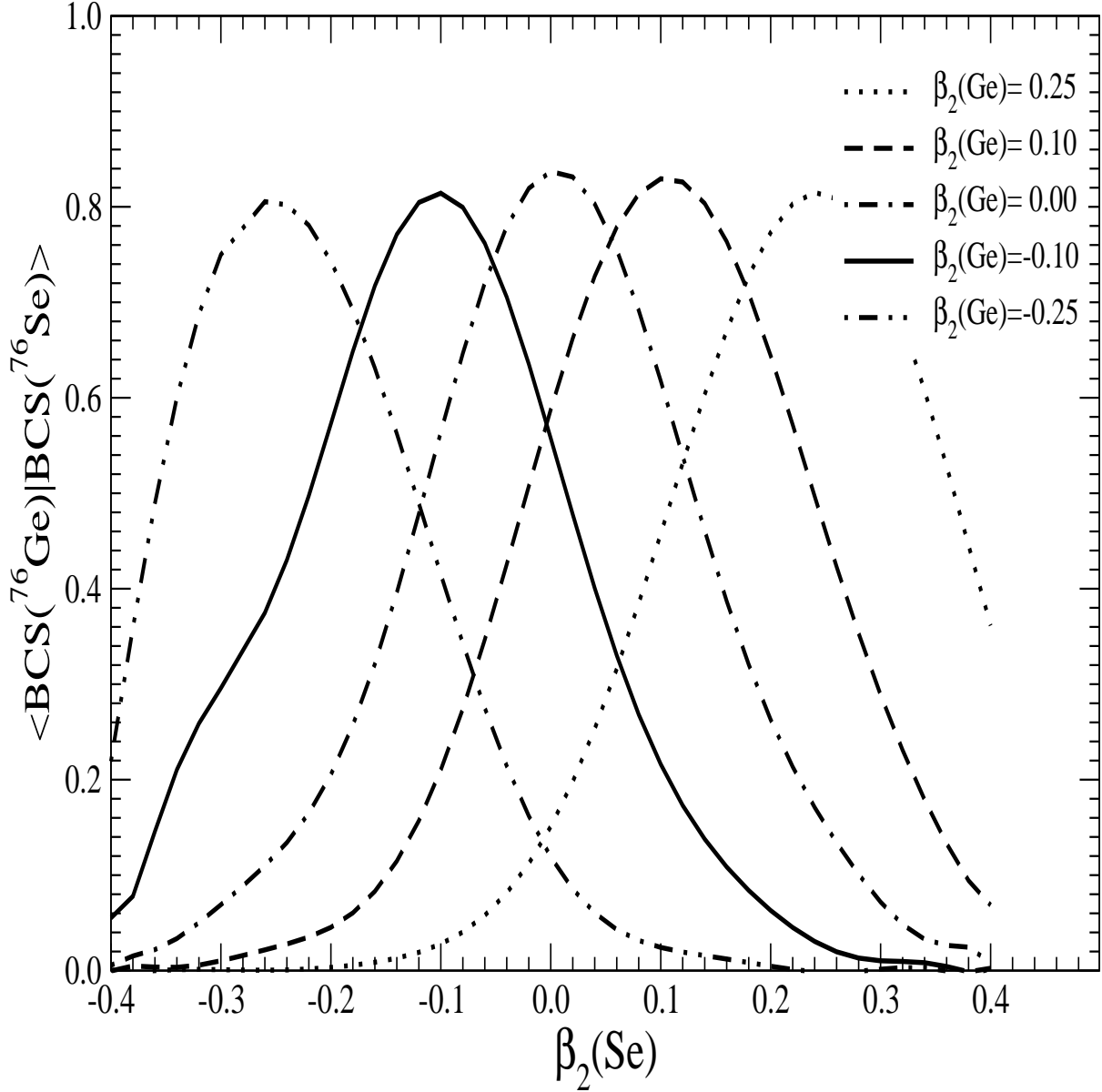}}
\caption{
The overlap factor of the initial and final BCS vacua as function
of the quadrupole deformation parameter $\beta_2$ of $^{76}Se$. 
The results are presented for spherical ($\beta_2=0.0$),
oblate ($\beta_2=-0.25, -0.10$) and prolate ($\beta_2=0.10, 0.25$) 
deformations of $^{76}Ge$.
}
\label{fig.5}
\end{figure}

One might ask what is the origin of this suppression. In Fig. 
\ref{fig.5} this point is clarified by presenting the overlap
factor of two BCS vacua for different values of $\beta_2(^{76}Ge)$
as function of the quadrupole deformation of $^{76}Se$. We see
that for a given $\beta_2(^{76}Ge)$ the curve has a maximum for
$\beta_2(^{76}Se)=\beta_2(^{76}Ge)$ and with increasing difference
in deformations of initial and final nuclei, i.e., $\Delta\beta_2$,
the value of the BCS overlap factor decreases rapidly. 
From the Fig. \ref{fig.5} it follows that oblate-prolate (or prolate-oblate) 
$2\nu\beta\beta$-decay transitions are disfavoured in comparison 
with prolate-prolate or oblate-oblate ones for the same 
absolute values of associated quadrupole deformation parameters.
In particular, by
assuming the $\beta_2$ and $\beta_4$ parameters of Ref. \cite{ring}
we obtain $<BCS(^{76}Ge)|BCS(^{76}Se)> = 0.009$. This suppression
is too strong to achieve agreement with experimental data for the
calculated $M_{GT}^{2\nu}$. However, by changing the deformation of 
$^{76}Se$ from oblate to prolate ($\beta_2(^{76}Se)=0.244$) we end up with
significantly larger value of the BCS overlap factor: 
$<BCS(^{76}Ge)|BCS(^{76}Se)> = 0.61$. 

\begin{figure}
\centerline{\psfig{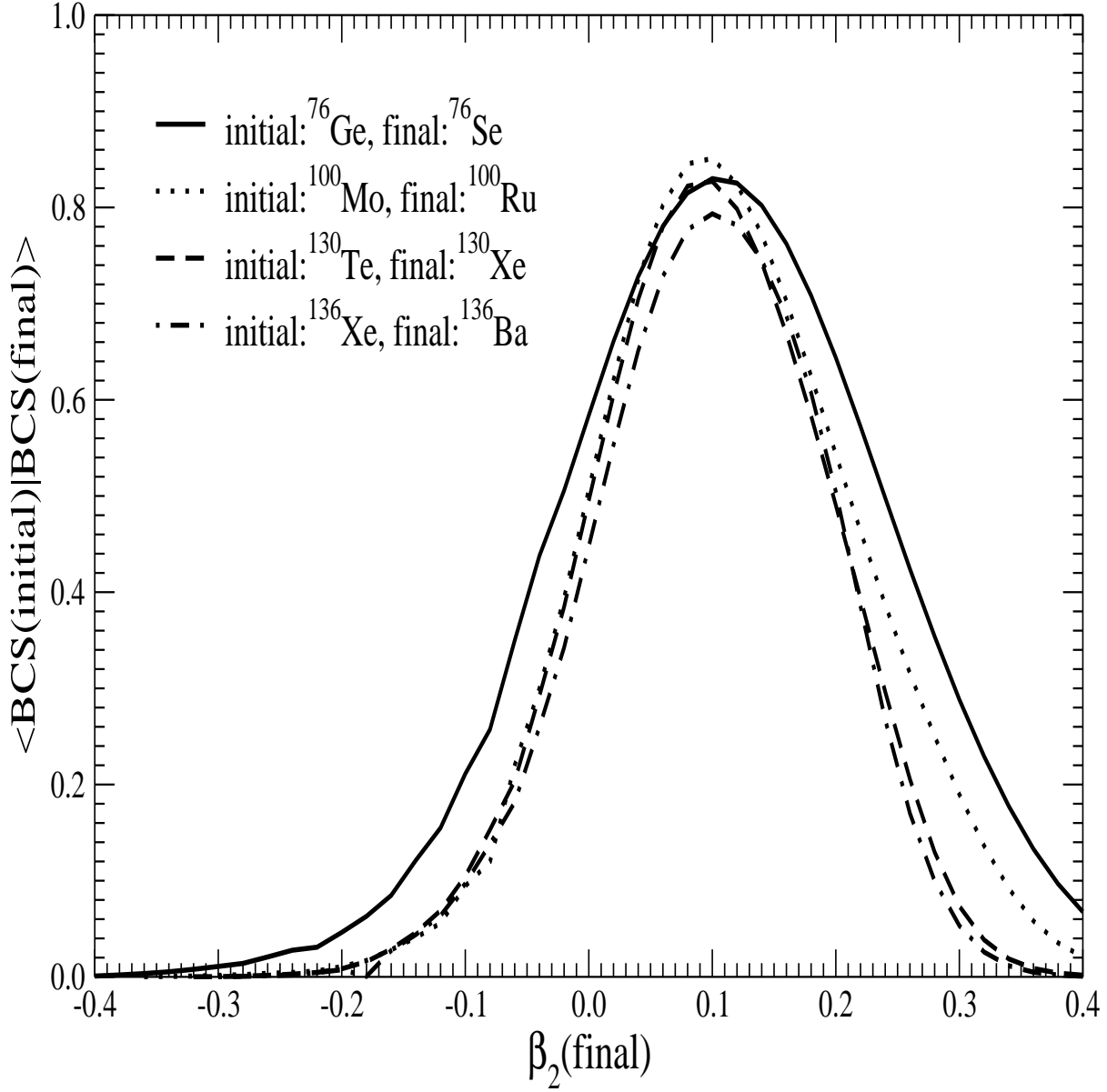}}
\caption{
The overlap factor of the initial and final BCS vacua as function
of the quadrupole deformation parameter $\beta_2$ of the final nucleus
for double beta decay of $^{76}Ge$, $^{100}Mo$, $^{130}Te$, and $^{136}Xe$.
The deformation parameter of the initial nucleus is chosen to be 
$\beta_2(initial)=0.1$.
}
\label{fig.6}
\end{figure}

The presented suppression mechanism is expecting to work also for other
$2\nu\beta\beta$-decay transitions. In Fig. \ref{fig.6} we present
BCS overlap factor for the $2\nu\beta\beta$-decay of $^{76}Ge$,
$^{100}Mo$, $^{130}Te$ and $^{136}Xe$. The quadrupole deformation
of the initial nucleus was taken to be $\beta_2=0.1$ and 
$\beta_2$ related with the final nucleus was considered to be a free
parameter  within the range 
$-0.4\le \beta_2 \le 0.4$. The pairing gaps, which 
entered the
BCS equations, are given in Eq. (\ref{eq:33}) for A=76 
and for A=100, 130 and 136 are given by
\begin{eqnarray}
^{100}Mo:\Delta_p = 1.612~MeV,~\Delta_n = 1.358~MeV, ~
^{100}Ru:\Delta_p = 1.548~MeV,~\Delta_n = 1.296~MeV,
\nonumber\\
^{130}Te:\Delta_p = 1.043~MeV,~\Delta_n = 1.180~MeV, ~
^{130}Xe:\Delta_p = 1.299~MeV,~\Delta_n = 1.243~MeV,
\nonumber\\
^{136}Xe:\Delta_p = 0.971~MeV,~\Delta_n = 1.408~MeV, ~
^{136}Ba:\Delta_p = 1.245~MeV,~\Delta_n = 1.032~MeV.
\end{eqnarray}
We see that the behaviour of the overlap factor for all considered
nuclear systems is qualitatively the same. The maximum of this factor
appears for equal quadrupole deformation of the initial and final
nuclei. With increasing value of $|\beta_2(initial)-\beta_2(final)|$
the value of the overlap factor strongly decreases. This fact implies 
that the deformation of nuclei plays an important role in the calculation
of the $2\nu\beta\beta$-decay transitions and should be known with
high reliability.   

\begin{figure}
\centerline{\psfig{figure=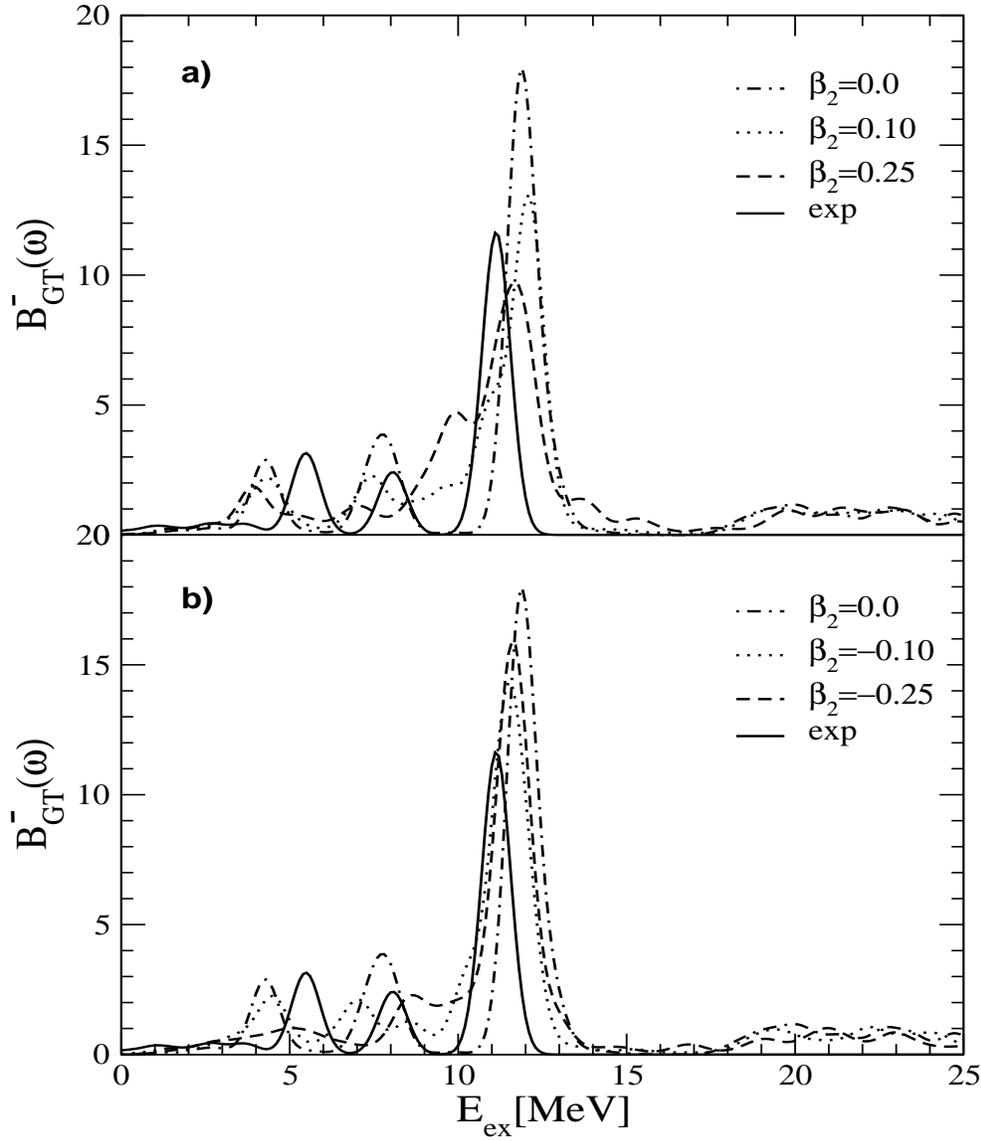,width=16cm,height=18cm,angle=0}}
\caption{
Gamov-Teller strength distributions $B^-_{GT}$  in $^{76}Ge$ 
as a function of the excitation energy of the daughter nucleus $E_{ex}$.
The distributions associated with prolate and oblate deformation of
$^{76}Ge$  are compared with that obtained for spherical shape
($\beta_2=0.0$)  in subfigures (a) and (b), respectively. 
The recommended values of $\chi$ and $\kappa$ of Ref. 
\protect\cite{homa} were considered: $\chi = 0.25~MeV$ and 
$\kappa = 0.028~MeV$. 
Experimental data (thick solid line) are from Ref. \protect\cite{madey}. 
}
\label{fig.7}
\end{figure}

\begin{figure}
\centerline{\psfig{figure=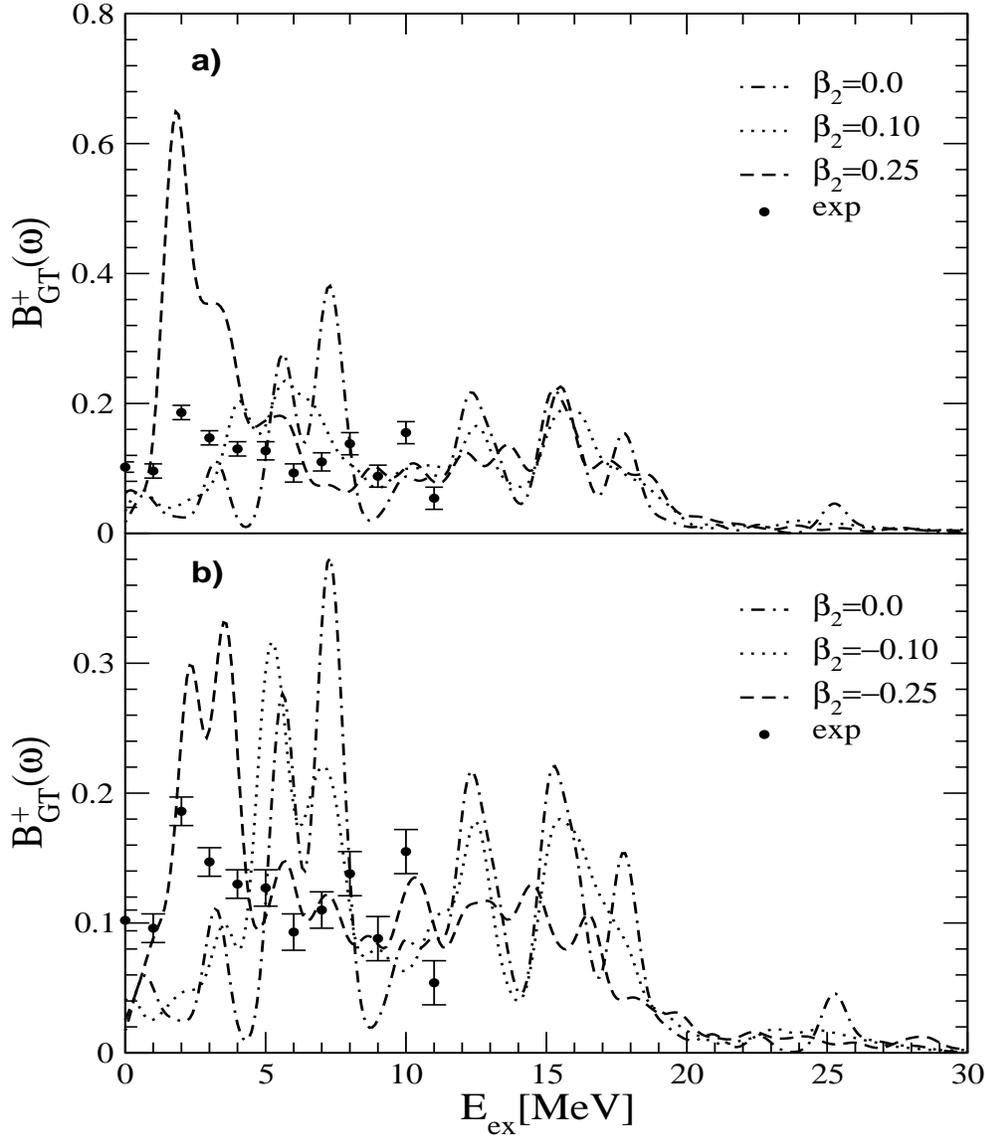,width=16cm,height=18cm,angle=0}}
\caption{
Gamov-Teller strength distributions $B^+_{GT}$  in $^{76}Se$ 
as a function of the excitation energy of the daughter nucleus $E_{ex}$.
The distributions associated with prolate and oblate deformation of
$^{76}Ge$  are compared with that obtained for spherical shape
($\beta_2=0.0$)  in subfigures (a) and (b), respectively. 
It is assumed $\chi = 0.25~MeV$ and $\kappa = 0.028~MeV$
\protect\cite{homa}. 
The data represented by the solid points are
from Ref. \protect\cite{helmer}.
}
\label{fig.8}
\end{figure}

An interesting issue is whether it is possible to obtain information 
about the deformations of $^{76}Ge$ and $^{76}Se$ by studying
the $\beta^-$  and $\beta^+$ strength distributions in these nuclei, 
respectively. For different values of quadrupole parameter $\beta_2$
they are presented in Figs. \ref{fig.7} and \ref{fig.8}. 
In order to facilitate comparison among various calculations the 
Gamow-Teller distributions are smoothed with a Gaussian of width $1~MeV$. 
The assumed strengths of the particle-hole and particle-particle interaction 
strength were $\chi = 0.25~MeV$ and $\kappa = 0.028~MeV$ \cite{homa}. 
By glancing at  Fig. \ref{fig.7}  we see
that the position of the Gamow-Teller resonance is not sensitive
to effect of deformation and that only a little strength is concentrated
above the region of the Gamow-Teller resonance. However,
there are significant differences between the spherical and
deformed strength distributions in general \cite{pedro}
especially in the case of prolate shape of $^{76}Ge$
(see Fig. \ref{fig.7}a), which is 
favoured by experimental data \cite{ragha}.
The thick solid line 
in Fig. \ref{fig.7} corresponds to data obtained from charge exchange 
$^{76}Ge(p,n)^{76}As$ reaction at $134.4~MeV$ \cite{madey}, which were 
folded with 1 MeV  width Gaussians as it has been done for the 
theoretical results. We see that the presented calculations are
in a very good agreement with data for the position 
of Gamow-Teller resonance.
It confirms the correct choice of the particle--hole strength
of the nuclear Hamiltonian ($\chi = 0.25~MeV$) in our calculation
\cite{homa}. 
From Fig. \ref{fig.7} it also follows that the strength in the peak 
of the experimental $B_{GT}^-$ distribution is better reproduced
by the  QRPA calculations with only sligtly deformed mean field.

The calculated $\beta^+$ strength distributions in $^{76}Se$ are 
displayed in Fig. \ref{fig.8}. They have been folded with
$\Gamma = 1~MeV$ width Gaussians.  We see that the effect of deformation
on the $\beta^+$ distribution strength for Gamow-Teller transition 
operator is more apparent as it is in the case  
$\beta^-$ strength distribution in $^{76}Ge$. In Fig. \ref{fig.8} we present also
the experimental data, which are represented by solid points. 
They were obtained from charge exchange $^{76}Se(n,p)^{76}As$ reaction
\cite{helmer}. We note that the experimentally  extracted strength distribution
with the help of two plausible methods differ from  each other
by more than a factor of 2 \cite{helmer}. Hence the experimental data
must be considered to be of a qualitative nature only.

There are two suppression mechanisms of the $2\nu\beta\beta$-decay 
matrix element. Already long ago Vogel and Zirnbauer showed that
$M^{2\nu}_{GT}$ is strongly suppressed when a reasonable  amount
of particle-particle interaction is taken into account
\cite{vogel2}, what is achieved close to a collapse of the 
QRPA solution. In this paper we show now that 
the $2\nu\beta\beta$-decay matrix element $M^{2\nu}_{GT}$ 
depends strongly on the difference in deformations of the initial and
final nuclei. A possible criterium to decide, which deformation for the 
initial and the final nucleus in the double beta decay should be used
could be the requirement for a common description of both single $\beta$ and 
$2\nu\beta\beta$-decay within the same nuclear Hamiltonian. 
In order to clarify this point three different
cases are considered:\\
i) Spherical shapes of both initial and final nuclei,
i.e., $\beta_2(^{76}Ge)=0$ and $\beta_2(^{76}Se)=0$.
\\
ii) Deformations of parent and daughter nuclei:
$\beta_2(^{76}Ge)=0.1$ and $\beta_2(^{76}Se)=0.266$.
\\
iii) Quadrupole deformations of the ground state of $^{76}Ge$ and 
$^{76}Se$: $\beta_2(^{76}Ge)=0.1$ and $\beta_2(^{76}Se)=0.216$.\\
The choice of the deformation parameters in cases i) and ii) 
do not contradict the data \cite{balraj,ragha,raman}.

\begin{figure}
\centerline{\psfig{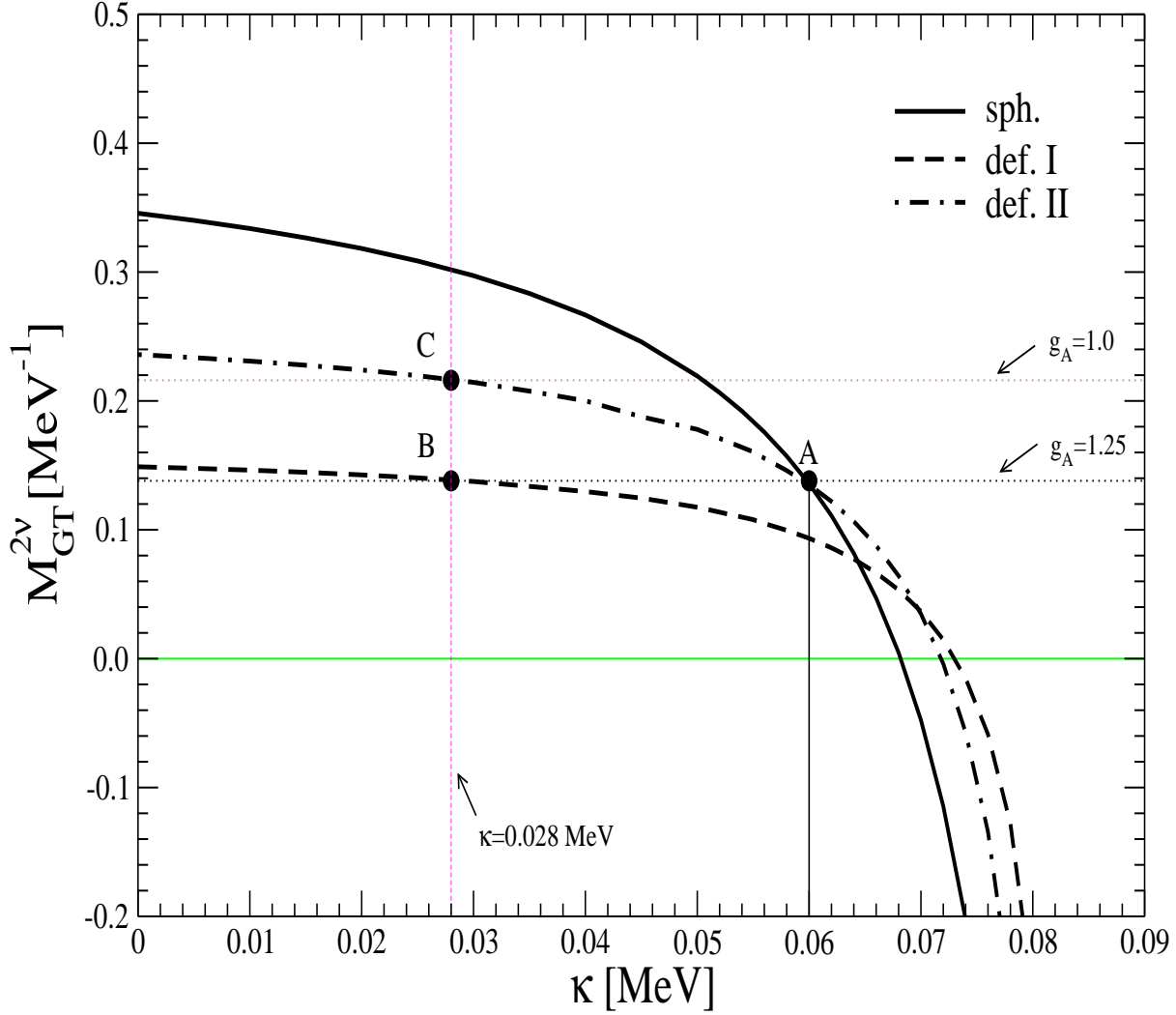}}
\caption{
$2\nu\beta\beta$-decay matrix element of $^{76}Ge$ as function of 
particle--particle interaction strength $\kappa$. 
The solid (sph.) line corresponds to spherical shape
of initial and final nuclei. 
The dashed (def. I) and dot-dashed (def. II) are asociated with 
a set of deformation parameters 
($\beta_2(^{76}Ge)=0.1$, $\beta_2(^{76}Se)=0.266$) and 
($\beta_2(^{76}Ge)=0.1$, $\beta_2(^{76}Se)=0.216$), respectively.
The points indicated by letters A, B and C determine $\kappa$ for which the 
value of $M_{GT}^{2\nu-exp.}$ deduced from the $2\nu\beta\beta$-decay 
half-life of $^{76}Ge$ is obtained. For $g_A = 1.25$ ($g_A = 1.0$)
one finds $M^{2\nu-exp}_{GT} = 0.138~MeV^{-1}$ 
($M^{2\nu-exp}_{GT} = 0.216~MeV^{-1}$) by assuming 
$T^{2\nu}_{1/2}(^{76}Ge) = 1.43\times 10^{21}$ years
\protect\cite{barab}.} 
\label{fig.9}
\end{figure}

In Fig. \ref{fig.9} the $2\nu\beta\beta$-decay of $^{76}Ge$ is presented as
function of particle-particle interaction strength $\kappa$ for the above
three cases. In the calculation determined by case i) (spherical nuclei)
$M_{GT}^{2\nu-exp.}$ equal to $0.138~MeV^{-1}$ ($g_A=1.25$) 
is reproduced for $\kappa = 0.06~MeV$. The corresponding set
of nuclear structure input parameters in this calculation is denoted by capital
letter A  (see Table \ref{table.1}). 
This value of $\kappa$ is significantly larger as the average value 
$\kappa_\beta = 0.028~MeV$ deduced from the systematic study of the single 
$\beta^+$-decays \cite{homa}. One can hardly suppose that it is because 
$^{76}Ge$ and $^{76}Se$ are specific nuclear systems. 
They are no indications in favor
of it. It seems to be a general problem within the spherical QRPA:
the calculated nuclear  matrix elements reproduce
the experimental $2\nu\beta\beta$-decay half-lifes as a rule 
for particle-particle strength  close 
to its critical value given by the QRPA collapse. 
In the case the calculation determined by case ii) the situation is different.
$M_{GT}^{2\nu-exp.}$ is reproduced for considerable smaller value of 
particle-particle strength $\kappa = 0.028~MeV$, which corresponds to 
that of Homa et al. \cite{homa}. The letter B donotes the coresponding
set of nuclear parameters, which is listed in Table \ref{table.1}.  
If we consider the effective axial coupling constant $g_A = 1$,
used often in the nuclei, $M_{GT}^{2\nu-exp.}$ is larger, namely $0.216~MeV^{-1}$.
Then, a  smaller difference between deformations of nuclei entering
$2\nu\beta\beta$-decay process is needed to reproduce this value for 
$\kappa \approx \kappa_\beta$. This is achieved with calculation 
of $M_{GT}$ performed within the case iii) for $\kappa = 0.028~MeV$
(nuclear parameter set C). Nevertheless, one should keep in mind that the 
coupling strength of \cite{homa} has been adjusted by using of a deformed Nilsson
single-particle model and $g_A = 1.25$. Thus it is not necessarily the best
possible choice. It is supposed that this prediction for $\kappa$ is more
accurate for nuclei with shorter $\beta$-decay half-lifes.
We note that that the curves corresponding to cases ii) and iii) do change
only weakly in the range $0\le \kappa \le \kappa_\beta $. It means that 
the role of the particle-particle interaction   within this interval is
negligible.

It is worthwhile to notice that for a large value of $\kappa$ 
of 0.07 MeV to 0.075 the agreement with $|M_{GT}^{2\nu-exp}|$
deduced from the  $2\nu\beta\beta$-decay half-life is achieved 
as well and that in this case the sign of $M_{GT}^{2\nu}$ is negative.
However, for negative value 
of $M_{GT}^{2\nu}$ the correspondance with $\kappa_\beta$ from 
systematic studies of the single beta decay \cite{homa} 
is not achieved. Thus negative values of $M_{GT}^{2\nu}$ 
are disfavored due to a complete disagreement with
single $\beta$-decay study of Homma et al. \cite{homa}.

In what follows we shall discuss the single $\beta$ decay characteristics
corresponding to three different  results in Fig. \ref{fig.4},
denoted by letters A, B and C, for which the experimental half-life
of the $2\nu\beta\beta$-decay of $^{76}Ge$ is reproduced.
The corresponding mean field and residual interaction  parameters 
are given in Table \ref{table.1}. 

\begin{figure}
\centerline{\psfig{figure=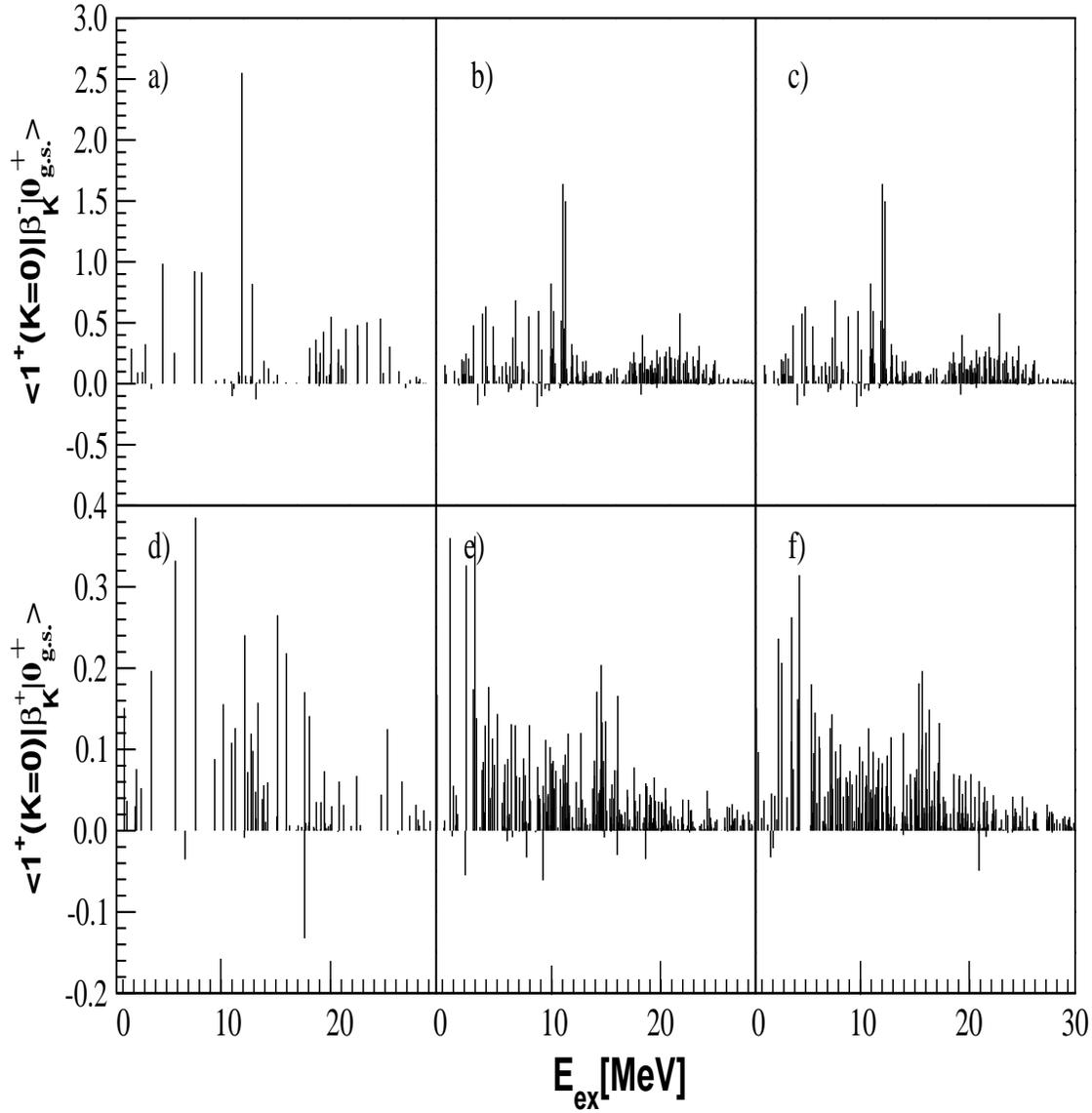,width=19cm,height=17cm,angle=-90}}
\caption{The $\beta^-$ ($\beta^+$) transition
amplitudes between the ground state
of $^{76}Ge$ ($^{76}Se$) and $K=0$ intermediate states
as function of the excitation energy $E_{ex}$ of $^{76}As$.
The $\beta^-$ [$\beta^+$] results obtained for input
parameter sets A, B, C 
(see Table \protect\ref{table.1}) are presented in subfigures a) [d)],
b) [e)] and c) [f)], respectively. Please observe, that the scale is 
different for the $\beta^-$ and the $\beta^+$ strength distributions.
}
\label{fig.10}
\end{figure}

\begin{figure}
\centerline{\psfig{figure=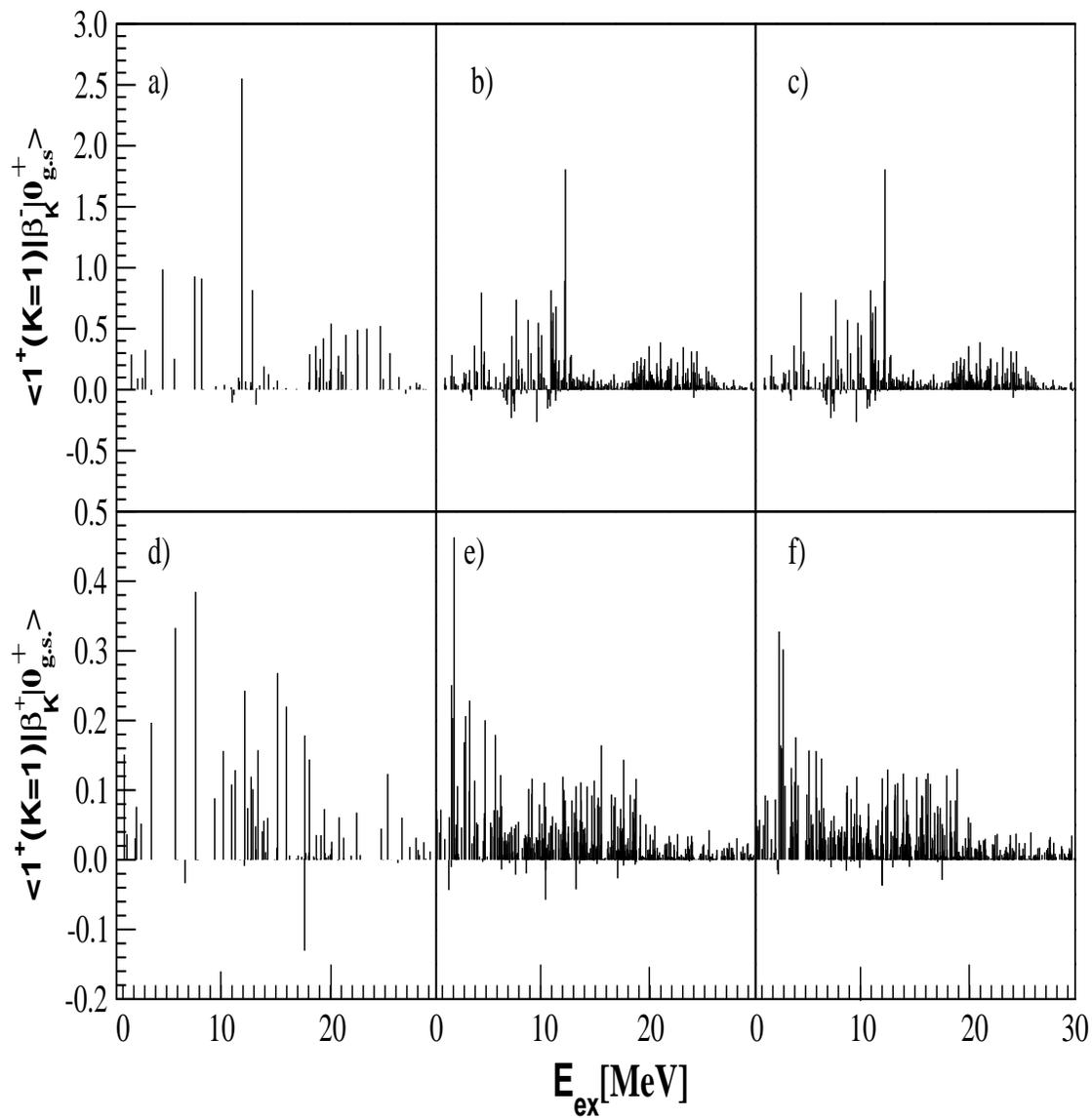,width=19cm,height=17cm,angle=-90}}
\caption{ The same as in Fig. \protect\ref{fig.10} for transitions to
$K=\pm 1$ intermediate states.
}
\label{fig.11}
\end{figure}

The $\beta^-$ and $\beta^+$ 
amplitudes calculated from the ground state of $^{76}Ge$ and
$^{76}Se$, respectively, associated with the $K=0$ ($K=\pm 1$) 
intermediate states  of $^{76}As$ are drawn in Fig. \ref{fig.10}
(Fig. \ref{fig.11}). By glancing these figures we see that deformation
(B and C sets of parameters) contribute to a fragmentation of the 
$\beta$-decay amplitudes and that values of $\beta^+$ amplitudes associated
with higher lying states of the intermediate nucleus (up to 20 MeV)
are significant. It is worthwhile  to note that 
the absolute values of $K=0$ and $K=\pm 1$ amplitudes are equal 
each to other in case of the QRPA calculation with spherical mean field.
The phases of these amplitudes are not of physical meaning. 
However, the relative phases of corresponding $\beta^+$ and $\beta^-$ 
amplitudes plays a crucial role due to partial destructive 
character of the contributions from different intermediate states to 
$M^{2\nu}_{GT}$. In the calculation these phases are fixed in such
way that diagonal elements of the overlap matrix in Eq. (\ref{eq:29}) 
are taken to be positive. 
We note that the $K=0$ and $K=\pm 1$ contributions to 
$M^{2\nu}_{GT}$ differ only slightly. 

\begin{figure}
\centerline{\psfig{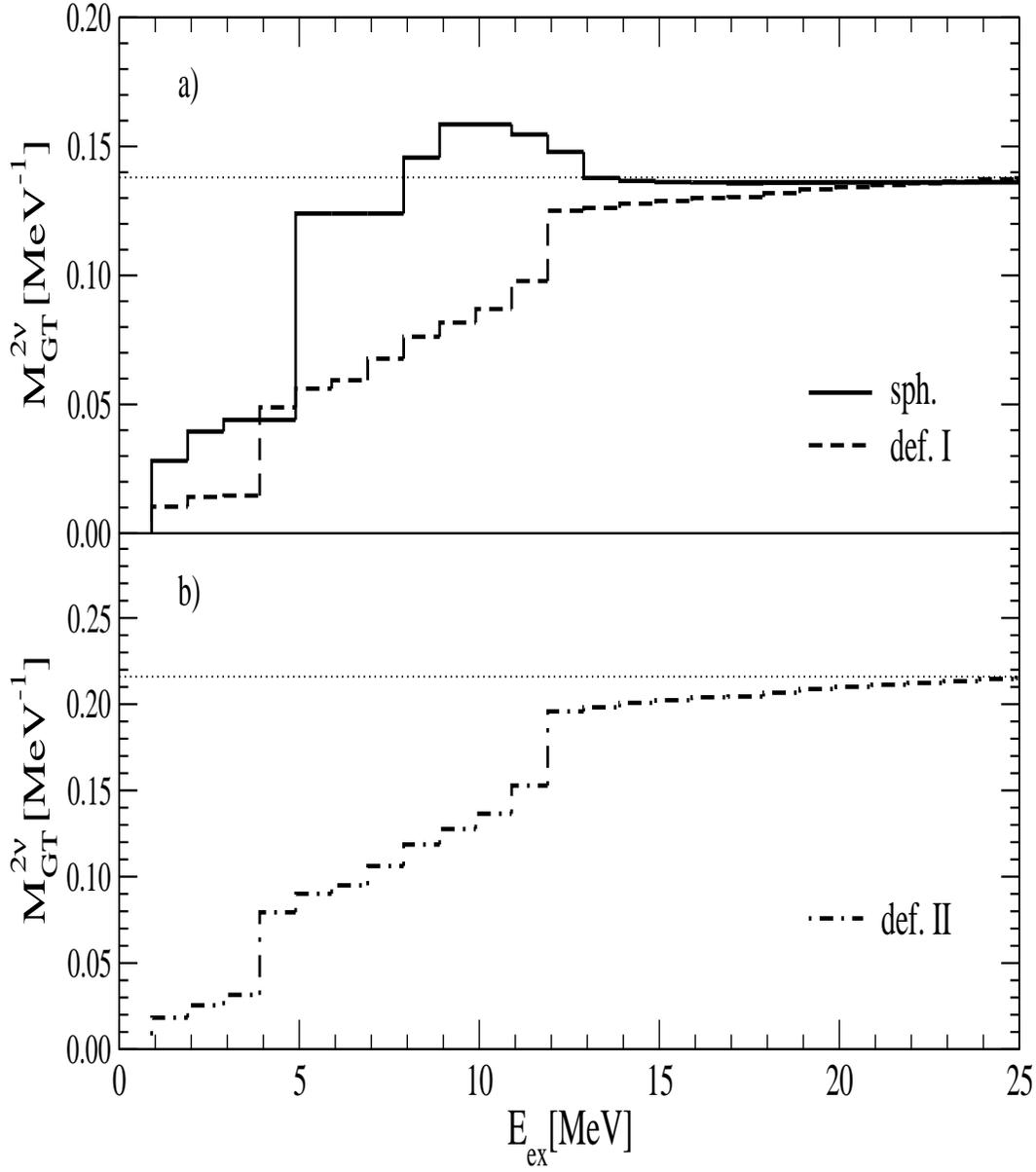}}
\caption{
Running sum of the $2\nu\beta\beta$-decay matrix element $M^{2\nu}_{GT}$
as a function of the excitation energy $E_{ex}$ in $^{76}As$. 
In the upper figure (a) results corresponding to input parameter sets 
A (solid line) and B
(dashed line) are presented. In the lower figure (b)  results obtained
with input parameter set C (dot-dashed line) are drawn. The 
dotted horizontal line in subfigure (a) [subfigure (b)] donotes the value of
matrix element $M^{2\nu-exp}_{GT} = 0.138~MeV^{-1}$ 
($M^{2\nu-exp}_{GT} = 0.216~MeV^{-1}$)  deduced from the
experimental $2\nu\beta\beta$-decay half-life of $^{76}Ge$
by assuming $g_A=1.25$ ($g_A=1.0$).
}
\label{fig.12}
\end{figure}

It is worthwhile to notice that a lot of $\beta^+$ strength in $^{76}Se$
is concentrated in the high energy region above $13~MeV$ 
(see Fig. \ref{fig.8}), 
which can not be described within the shell model.  There is a 
longstanding question whether in the calculation of the 
$2\nu\beta\beta$-decay matrix element $M^{2\nu}_{GT}$ the transitions
to higher lying states of the intermediate nucleus play an important 
role. There is so called ``Single State Dominance Hypothesis'' (SSDH), 
which assumes that only the lowest $1^+$ state of the intermediate nucleus
is of major importance in the evaluation of $M^{2\nu}_{GT}$. 
The SSDH can be realized, e.g., through cancellation among the higher
lying $1^+$ states of the intermediate nucleus. In Ref. \cite{ssd}
a possibility to decide experimentally  about validity of the SSDH 
by measuring the single electron spectra and the angular distribution
of the emitted electrons was discussed. This treatment of the SSDH
is suitable for the $2\nu\beta\beta$-decay transitions with low-lying
$1^+$ ground state of the intermediate nucleus. But this is not the case
of the $2\nu\beta\beta$-decay of $^{76}Ge$. In Fig. \ref{fig.12} the
QRPA model dependent study of this problem is presented by drawing
the running sum of matrix element $M^{2\nu}_{GT}(E)$ as a function
of the $1^+$ excitation energy $E_{ex}$ of the intermediate nucleus.
We see that in the case of the spherical calculation the main contribution
to $M^{2\nu}_{GT}$ comes from the states of the intermediate nucleus
lying below 5 MeV scale and that there is a partial cancellation among
contributions to $M^{2\nu}_{GT}$ from higher lying states. In the case of 
deformed QRPA calculations there is a different situation. 
We see that important contributions to  $M^{2\nu}_{GT}(E)$ arise even from 
relatively high energies of the intermediate states around the position of 
Gamow-Teller resonance (about 12 MeV), which can not
be described for a given nuclear system in the framework of the
shell model. These results contradict the SSDH 
and indicate that a strong truncation in consideration of the complete set
of the states of the intermediate nucleus is not appropriate.

\section{Summary and Conclusions}

For the $2\nu\beta\beta$-decay ground state 
transition $^{76}Ge \rightarrow {^{76}Se}$ a detailed study of
nuclear matrix elements within the deformed QRPA with separable 
spin--isospin interactions in the particle--hole and particle--particle 
channels was performed. A new mechanism of the suppression of the 
double beta decay nuclear matrix elements was found. 
We showed that the effect of deformation on the $2\nu\beta\beta$-decay
matrix element is large for a significant difference in deformations of the
parent and daughter nuclei and is not related with the increasing amount 
of the ground state correlations close to a collapse of the 
QRPA solution. The origin of this new mechanism is a strong sensitivity
of the overlap of the initial and final BCS vacua to the deformations
of the initial and final nuclei. This enters directly into the
the overlap  of the intermediate nuclear 
states generated from the initial and final nuclei via QRPA
diagonalization. A study of other double beta decay transitions as a
function of this deformation dependent overlap indicates the 
general importance of  the new suppression mechanism for the  
$2\nu\beta\beta$-decay and  $0\nu\beta\beta$-decay matrix elements.

Next, we showed that by assuming spherical nuclei there is a 
strong disagreement between the particle-particle strength
$\kappa$ needed to reproduce the  
$2\nu\beta\beta$-decay half-life of $^{76}Ge$ and 
the $\kappa$ deduced in Ref. \cite{homa} from a systematic study of the single
$\beta$-decay transitions. By the new suppression
mechanism of the $2\nu\beta\beta$-decay matrix
element we showed that for deformations of the parent 
and daughter 
nuclei, which do not contradict  the experimental data,  
a simultaneous description
of both single $\beta$ and $2\nu\beta\beta$-decay is possible. 
We also discussed the role of the axial coupling constant
$g_A$.

We found that a significant part of the total
$\beta^+$ strength is concentrated in the energy region above $10~MeV$.
A detail study of contributions to 
$M^{2\nu}_{GT}$ from different $1^+$ intermediate nuclear states 
showed that for the A=76 nuclei it is necessary to consider 
all $1^+$ states of $^{76}As$ up to energy $15-20$ MeV.  This fact disfavors
theoretical studies of the $2\nu\beta\beta$-decay of medium and
heavy nuclei in models with a strongly truncated basis, like the shell model.

The Gamow-Teller $\beta^+$ strengths and double beta decay matrix elements in 
medium and heavy nuclei continue to be a challenge for nuclear structure
models. Further theoretical studies are needed to clarify the role of
deformation in calculation of other double beta decay transitions, 
especially those including  heavier nuclear systems, which are known to be
strongly deformed.  Future works should investigate 
the $2\nu\beta\beta$-decay matrix elements also within different
possible extensions of the deformed QRPA, e.g., those including 
proton-neutron pairing \cite{pnp}. In this way one expects 
to develop a reliable many-body approach with well defined nuclear structure
parameters for calculation of the $0\nu\beta\beta$-decay 
matrix elements. Their accurate values are highly required to
determine the neutrino mixing pattern and to answer the question
\cite{bil99} about the dominant mechanism of the  $0\nu\beta\beta$-decay
\cite{erice}.

\section{Appendix A}

The single-particle states are calculated by solving the 
Schr\"odinger equation with the deformed axially symmetric Woods-Saxon 
potential, which parameterization is given in Ref. \cite{Tan79}.
They are characterized by their energy 
$\varepsilon_{\tau}$, parity $\pi_\tau$  and by the projection 
$\Omega_\tau$ ($\tau = p, n$) 
of the full angular momentum on the nuclear symmetry axis. We
use notation $|p \rho_p>$ and $|n \rho_n>$ for protons and neutrons, 
respectively. $|\tau \rho_\tau>$ represents proton ($\tau= p$) or 
neutron ($\tau = n$) state with quantum numbers $\Omega_\tau$
and $\pi_\tau$. $\rho_\tau$ is the sign of the angular momentum 
projection $\Omega$ ($\rho_\tau = \pm 1$).
We note that intrinsic states are twofold degenerate. The
states with $\Omega_\tau$ and $-\Omega_\tau$ have the same energy as
consequence of the time reversal invariance. We shall use notation
such that $\rho_\tau$ is taken to be positive for states and 
negative for time reversal states. 

In order to solve the Schr\"odinger equation the eigenfunctions
of a deformed symmetric harmonic oscillator are used as a basis for 
the diagonalization of the mean-field Hamiltonian \cite{Da135}. 
These states are completely determined by a principal set of 
quantum numbers $(N, n_z, \Lambda, \Omega)$, where 
$N=n_x+n_z$, $n_x=2n_r+|\Lambda|$ and
$\Omega = \Lambda + \Sigma$. $n_z-1$ and $n_x-1$ are number of nodes
of the basis functions in the $z$-direction and $r$-direction,
respectively. $\Lambda$ and $\Sigma$ are the
projections of the orbital and spin angular momentum on the
symmetry axis z. 
The explicit form of single-particle harmonic oscillator 
wave functions in cylindrical coordinates
$(r,z,\phi )$ can be found, e.g., in Ref. \cite{No324}.  
For a given shape of the nuclear surface, the shape of the 
deformed harmonic oscillator is automatically chosen in a way suitable 
for obtaining good accuracy with a smallest number of basis functions
\cite{Da135}. The deformation dependent cutt-off
is chosen in such way as to 
assure numerical stability of the results. In our calculation
we use 11 major shells.

In cylindrical coordinates the eigenfunctions of states 
and time-reversed states in
deformed Woods-Saxon potential are expressed as follows:
\begin{eqnarray}
|\tau \rho_\tau = +1> &=& \sum_{N n_z}
[ b^{(+)}_{N n_z \Omega_\tau} 
|N, n_z, \Lambda_\tau, \Omega_\tau=\Lambda_\tau+1/2> +
\nonumber \\
&&~~~~~
b^{(-)}_{N n_z \Omega_\tau}  
|N, n_z, \Lambda_\tau+1, \Omega_\tau=\Lambda_\tau+1-1/2>]
\end{eqnarray}
and
\begin{eqnarray}
|\tilde{\tau} \rho_\tau = +1> &=& |\tau \rho_\tau = -1> =
\sum_{N n_z}
[ b^{(+)}_{N n_z \Omega_\tau} 
|N, n_z, -\Lambda_\tau, \Omega=-\Lambda_\tau-1/2> -
\nonumber \\
&&~~~~~
 b^{(-)}_{N n_z \Omega_\tau} 
|N, n_z, -\Lambda_\tau-1, \Omega_\tau=-\Lambda_\tau-1+1/2>]
\end{eqnarray}
with $\Lambda \ge 0$. $\sim$ indicates time reversal states.

The single-particle matrix elements of the $\tau^+ \sigma_K$
operator are given by
\begin{eqnarray}
<p \rho_p|\tau^+ \sigma_{K=0}|n \rho_n> &=&
\delta_{\Omega_p \Omega_n} \rho_p \sum_{N n_z}
[ b^{(+)}_{N n_z \Omega_p} b^{(+)}_{N n_z \Omega_n}-
 b^{(-)}_{N n_z \Omega_p} b^{(-)}_{N n_z \Omega_n}],
\\
<p \rho_p|\tau^+ \sigma_{K=+1}|n \rho_n> &=& 
-\sqrt{2}\delta_{\Omega_p \Omega_n+1}
 \sum_{N n_z} b^{(+)}_{N n_z \Omega_p} b^{(-)}_{N n_z \Omega_n} 
~~~for~~ \rho_p = \rho_n = +1, \nonumber \\
&=&+\sqrt{2}\delta_{\Omega_p \Omega_n+1}
\sum_{N n_z} b^{(-)}_{N n_z \Omega_p} b^{(+)}_{N n_z \Omega_n} 
~~~for~~ \rho_p = \rho_n = -1, \nonumber \\
&=&-\sqrt{2}\delta_{\Omega_p \frac{1}{2}} \delta_{\Omega_n -\frac{1}{2}}
 \sum_{N n_z} b^{(+)}_{N n_z \Omega_p} b^{(+)}_{N n_z \Omega_n} 
~~~for~~ \rho_p = +1,~ \rho_n = -1,\\
<p \rho_p|\tau^+ \sigma_{K=-1}|n \rho_n> &=& 
 \sqrt{2}\delta_{\Omega_p \Omega_n-1}
 \sum_{N n_z} b^{(-)}_{N n_z \Omega_p} b^{(+)}_{N n_z \Omega_n} 
~~~for~~ \rho_p = \rho_n = +1, \nonumber \\
&=&-\sqrt{2}\delta_{\Omega_p \Omega_n-1}
\sum_{N n_z} b^{(+)}_{N n_z \Omega_p} b^{(-)}_{N n_z \Omega_n} 
~~~for~~ \rho_p = \rho_n = -1, \nonumber \\
&=&\sqrt{2}\delta_{\Omega_p -\frac{1}{2}} \delta_{\Omega_n \frac{1}{2}}
 \sum_{N n_z} b^{(+)}_{N n_z \Omega_p} b^{(+)}_{N n_z \Omega_n} 
~~~for~~ \rho_p = +1,~ \rho_n = -1.
\end{eqnarray}

The overlap of the proton and neutron single particle states 
of the initial $(A,Z)$ and final $(A,Z+2)$ nuclei are calculated
by assuming that the corresponding sets of basis wave functions
do not differ significantly each from other. Then we have
\begin{eqnarray}
<p_f \rho_{p_f}|p_i \rho_{p_i}> &=& 
\delta_{\Omega_{p_f} \Omega_{p_i}} 
\sum_{N n_z}
[ b^{(+)}_{N n_z \Omega_{p_f}} b^{(+)}_{N n_z \Omega_{p_i}}-
 b^{(-)}_{N n_z \Omega_{p_f}} b^{(-)}_{N n_z \Omega_{p_i}}],
\nonumber \\
<n_f \rho_{n_f}|n_i \rho_{n_i}> &=&
\delta_{\Omega_{n_f} \Omega_{n_i}} 
\sum_{N n_z}
[ b^{(+)}_{N n_z \Omega_{n_f}} b^{(+)}_{N n_z \Omega_{n_i}}-
 b^{(-)}_{N n_z \Omega_{n_f}} b^{(-)}_{N n_z \Omega_{n_i}}].
\end{eqnarray}
The index i (f) denotes that proton and neutron single particle 
states are defined with respect to the initial (final) nucleus.

\section{Appendix B}

As a consequence of considered many-body approximations
the two sets of intermediate nuclear states generated from the 
initial and final ground states are not identical within the
QRPA theory. Thus it is necessary to introduce the overlap
factor of these states, which can be expressed with the help
of intrinsic phonon operators as follows:
\begin{equation}
<1(K),m_f|1(K),m_i> = <rpa_f|~ Q^{m_f}_K ~ {Q^{m_i \dagger}_K}~|rpa_i>.
\label{eq:b1}
\end{equation}
Here, the index i (f) indicates that the excited states of the nucleus 
are defined with respect to the initial (final) nuclear ground state 
$|rpa_i>$ ($|rpa_f>$). 

In order to evaluate  $<1(K),m_f|1(K),m_i>$ we express the phonon
creation operator ${Q^{m_i}_K}^\dagger$ in terms of creation and
anninhilation phonon operators  associated with the final nucleus.
We have 
\begin{equation}
{Q^{m_i \dagger}_K} = \sum_{m_f}~ 
(a_{m_i m_f} 
{Q^{m_f \dagger}_K} ~ +~ b_{m_i m_f} {\tilde Q}^{m_f}_{K} ).
\label{eq:b2}
\end{equation}
The coefficients of expansion $a_{m_i m_f}$ and $b_{m_i m_f}$ will
be determined below. By inserting Eq. (\ref{eq:b2}) into 
Eq. (\ref{eq:b1}) we get
\begin{eqnarray}
<1(K),m_f|1(K),m_i> &=& \sum_{{m'}_f} [
<rpa_f| Q^{m_f}_K {Q^{{m'}_f \dagger}_K}|rpa_i> a_{m_i {m'}_f} 
 \nonumber \\
&& +<rpa_f| Q^{m_f}_K {Q^{{m'}_f}_{-K}}|rpa_i> b_{m_i {m'}_f} ]
\nonumber \\
&\approx&  a_{m_i {m'}_f} <BCS_f|BCS_i>.
\label{eq:b3}
\end{eqnarray}
Here, we neglected  overlap matrix element between the final ground state 
and the two-phonon state generated from the initial nucleus, which is 
considered to be small and should be not related with the studied
quantity. In addition, we approximated 
the overlap of initial and final RPA ground state with the BCS ones. 
Thus the overlap factor of the intermediate nuclear states
generated from the initial and final ground states is proportional to the
overlap of initial and final BCS vacua. 

We proceed with the calculation of $a_{m_i {m'}_f}$ and $b_{m_i {m'}_f}$
coefficients. The quasiparticle creation and anninhilation operators
$(a^{(i)\dagger},a^{(i)})$ [$(a^{(f)\dagger},a^{(f)})$] of the initial 
[final] nucleus are connected with the particle creation and  annihilation 
operators  $(c^{(i)\dagger},c^{(i)})$ [$(c^{(f)\dagger},c^{(f)})$] 
by the BCS transformation [see Eq. (\ref{eq:4})]. In addition there is a 
unitary transformation between particle operators associated with 
initial and final nuclei
\begin{eqnarray}
{c^{(i) \dagger}_{\tau \rho_\tau}} &=& \sum_{\tau'{\rho}_{\tau'}} 
<\tau \rho_\tau |\tau' {\rho}_{\tau'} > {c^{(f) \dagger}_{\tau' {{\rho}_{\tau'}}}}
\nonumber\\
{{\tilde c}^{(i)}_{\tau \rho_\tau}} &=& \sum_{\tau'{\rho}_{\tau'}} 
<\tau \rho_\tau |\tau' {\rho}_{\tau'} > {{\tilde c}^{(f)}_{\tau' {{\rho}_{\tau'}}}} 
\label{eq:b4} 
\end{eqnarray}
The overlap factors of the single particle wave functions
of the initial and final nuclei $<\tau \rho_\tau |\tau' {\rho'}_{\tau'} >$ 
is given explicitely in the Appendix A. The above mentioned transformations allow
us, by using the quasiboson approximation, to rewrite the boson
operators of the initial nucleus with the help of the boson operators 
of the final nucleus:
\begin{eqnarray}
A^{(i) \dagger} (l,K) &=& \sum_{l'} [ {\cal R}_{ll'} A^{(f) \dagger} (l',K) 
+ {\cal S}_{ll'} A^{(f)} ({\overline l}',K) ], \nonumber \\
A^{(i)} ({\overline l},K) &=& \sum_{l'} [ {\cal R}_{ll'} A^{(f)} ({\overline l}',K) 
- {\cal S}_{ll'} A^{(f) \dagger} ({l}',K) ]. 
\label{eq:b5} 
\end{eqnarray}
Assuming the definition of the quasiparticle pairs operator  in Eq. (\ref{eq:11})
the factors ${\cal R}_{ll'}$ and  ${\cal S}_{ll'}$ can be expressed as 
\begin{eqnarray}
{\cal R}_{ll'}&=& <p \rho_p |p' {\rho}_{p'} >(u^{(i)}_p u^{(f)}_{p'}+v^{(i)}_p v^{(f)}_{p'}) 
 <n \rho_n |n' {\rho}_{n'} >(u^{(i)}_n u^{(f)}_{n'}+v^{(i)}_n v^{(f)}_{n'}), 
\nonumber\\
{\cal S}_{ll'}&=& <p \rho_p |p' {\rho}_{p'} >(u^{(i)}_p v^{(f)}_{p'}-u^{(i)}_p v^{(f)}_{p'}) 
 <n \rho_n |n' {\rho}_{n'} >(u^{(i)}_n v^{(f)}_{n'}-u^{(i)}_n v^{(f)}_{n'}).
\label{eq:b6} 
\end{eqnarray}
It is worthwhile to notice that in the limit initial and final states are indentical
${\cal R}_{ll'}=1$ and  ${\cal S}_{ll'}=0$. 

By inserting Eq. (\ref{eq:b5}) into the expression for the phonon operator of the
initial nucleus [see Eq. (\ref{eq:13})]
and by exploiting the relations
\begin{eqnarray}
A^{(f) \dagger} (l, K) &=& 
\sum_{m_f} [ X^{m_f}_{i, K} {Q^{m_f \dagger}_K} + 
Y^{m}_{i, K} {{\tilde Q}^m_K}],\nonumber\\
A^{(f)} ({\overline l}, K) &=& 
\sum_{m_f} [ X^{m_f}_{i, K} {{\tilde Q}^{m_f}_K} + 
Y^{m_f}_{i, K} {{Q}^{m_f \dagger}_K}],
\label{eq:b7}
\end{eqnarray}
we find \cite{pan88}
\begin{eqnarray}
a_{m_i m_f} &=& \sum_{l l'}
[ X^{m_f}_{l' K} {\cal R}_{l' l} X^{m_i}_{l K}-
  Y^{m_f}_{l' K} {\cal R}_{l' l} Y^{m_i}_{l K} \nonumber\\
&& ~+~ 
  Y^{m_f}_{l' K} {\cal S}_{l' l} X^{m_i}_{l K}-
  Y^{m_f}_{l' K} {\cal S}_{l' l} Y^{m_i}_{l K} ].
\label{eq:b8}
\end{eqnarray}
By neglecting the terms proportional to ${\cal S}_{ll'}$ due to their
smallness we end up with the overlap factor of the intermediate nuclear 
states given in Eq. (\ref{eq:29}).

The overlap factor of the initial and final BCS vacua 
can be written as product of proton and neutron
BCS overlap factors for a given anular momentum projection
quantum number $\Omega$:
\begin{eqnarray}
<BCS_f|BCS_i> &=& <BCS_f(p)|BCS_i(p)> <BCS_f(n)|BCS_i(n)> 
\nonumber \\
&=& \prod_{\Omega_p}<BCS_f(\Omega_p )|BCS_i(\Omega_p )> 
\prod_{\Omega_n}<BCS_f(\Omega_n )|BCS_i(\Omega_n )>. 
\label{eq:b9}
\end{eqnarray}
where
\begin{equation}
<BCS_f(\Omega )|BCS_i(\Omega )>~ =~ 
<| \prod_{k=1}^{N_\Omega} \left(u^{(f)}_{k} + v^{(f)}_{k}
c^{(f)}_{\overline k}  c^{(f)}_{k}\right)~
\prod_{l=1}^{N_\Omega} \left(u^{(i)}_{l} + v^{(i)}_{l}
 c^{(i) \dagger}_{l} c^{(i) \dagger}_{\overline l}
\right) |>.
\label{eq:b10}
\end{equation} 
$N_\Omega$ is the number of single particle states with the same value
of quantum number $\Omega$. The same model space 
for protons and neutrons is assumed.
By a direct calculation of the above matrix element one finds
\begin{eqnarray}
&&<BCS_f(\Omega )|BCS_i(\Omega )> = 
\prod_{k=1}^{N_\Omega} u^{(f)}_{k}~ \prod_{l=1}^{N_\Omega} u^{(f)}_{l}
\nonumber \\
&+& \sum_{m_1,n_1=1}^{N_\Omega}~ 
v^{(f)}_{{m_1}}~ v^{(i)}_{{n_1}}~ 
\left( D^{(1)}(m_1;n_1)\right)^2~
\prod_{k=1}^{N_\Omega (m_1)} u^{(f)}_{k}~
\prod_{l=1}^{N_\Omega (n_1)} u^{(i)}_{l}
\nonumber \\
&+& \sum_{m_1,m_2,n_1,n_2=1}^{N_\Omega}~ 
v^{(f)}_{{m_1}} v^{(f)}_{{m_2}}~ 
v^{(i)}_{{n_1}} v^{(i)}_{{n_2}}
 \left( D^{(2)}(m_1,m_2;n_1,n_2)\right)^2~
\prod_{k=1}^{N_\Omega (m_1,m_2)} u^{(f)}_{k}~
\prod_{l=1}^{N_\Omega (n_1,n_2)} u^{(i)}_{l}
\nonumber\\
&+&~\cdot\cdot\cdot~+~
\left( D^{(N_\Omega)}(1,2,...,N_\Omega;1,2,..,N_\Omega )\right)^2~
\prod_{k=1}^{N_\Omega } v^{(f)}_{{k}} ~
\prod_{l=1}^{N_\Omega } v^{(i)}_{{l}}.
\label{eq:b11}
\end{eqnarray} 
Here, $\prod_{k=1}^{N_\Omega (m_1,m_2)}$ means that index $k$ runs
the values from 1 to $N_\Omega$ except the values $k=m_1$ and $k=m_2$
($1 \le m_1 \le N_\Omega$ and $1 \le m_2 \le N_\Omega$). 
$D^{r}(m_1,m_2,...,m_r;n_1,n_2,...,n_r)$  denotes the determinant of matrix of
rank $r$ constructed of elements of the unitary matrix of the transformation
between the initial and final single particle states with row indices
$m_1, m_2, ...,m_r$ and column indices $n_1, n_2, ...,n_r$. 
It is worthwhile to notice that by replacing all determinants 
in Eq. (\ref{eq:b11}) with
unity, i.e. the matrix of the transformation between the single
particles associated with both nuclei is just unity matrix, we obtain
a compact expression \cite{grotz85}
\begin{equation}
<BCS_f(\Omega )|BCS_i(\Omega )> = 
\prod_{k=1}^{N_\Omega} (u^{(f)}_{k} u^{(i)}_{k} + v^{(f)}_{k} v^{(i)}_{k} ).
\end{equation}
However, this approximation is not justified and can lead to a
significant inaccuracy in the calculation of $M_{GT}^{2\nu}$ especially
if there is a strong difference in deformations of the intial and final nuclei.

\section{Acknowledgements}

We are thankful to A.A. Raduta and P. Sarriguren for 
stimulating comments and discussions. 
This work was supported in part by the Deutsche 
Forschungsgemeinschaft (436 SLK 17/298), by the
``Land Baden-W\"urtemberg'' as a ``Landesforschungsschwerpunkt:
Low Energy Neutrinos'' and by the 
VEGA Grant agency of the Slovac Republic under contract
No. 1/0249/03.

\newpage


\begin{table}[t]
\caption{Three sets of nuclear structure input 
parameters (A, B, and C) for which the 
calculated nuclear matrix elements $M^{2\nu}_{GT}$ 
reproduces the experimental
$2\nu\beta\beta$-decay half-life of $^{76}Ge$.
v.o.f. denotes overlap factor of the initial and final BCS vacua.}
\label{table.1}
\begin{tabular}{ccccccccccccc}\hline
 & \multicolumn{3}{c}{mean field of $^{76}Ge$} & &
   \multicolumn{3}{c}{mean field of $^{76}Se$} & & & & &
\\ \cline{2-4} \cline{6-8} 
 & \multicolumn{1}{c}{Def.} &  \multicolumn{2}{c}{Pairing} &  
 & \multicolumn{1}{c}{Def.} &  \multicolumn{2}{c}{Pairing}   
 & & \multicolumn{2}{c}{$H_{int}$} & \\
\cline{2-4} \cline{6-8}  \cline{10-11}
par. & $\beta_2$ & $\Delta_p$ & $\Delta_n$ & &
$\beta_2$  & $\Delta_p$ & $\Delta_n$ & v.o.f. &
$\chi$ & $\kappa$ & $g_A$ & $M^{2\nu}_{GT}$ \\
 set  & &  $MeV$ & $MeV$ &  & &  $MeV$ & $MeV$ & & $MeV$ & $MeV$ &
 & $MeV^{-1}$ \\ \hline
A & 0.0   &   1.561 & 1.535 & &
    0.0   &   1.751 & 1.710 & 0.842 & 0.25 & 0.060 & 1.25 & 0.138\\
B & 0.10 &   1.561 & 1.535 & &
    0.266 &   1.751 & 1.710 & 0.403 & 0.25 & 0.028 & 1.25 & 0.138\\
C & 0.10 &   1.561 & 1.535 & &
    0.216 &   1.751 & 1.710 & 0.587  & 0.25 & 0.028 & 1.00 & 0.216\\
\hline
\end{tabular}
\end{table}


\newpage

\newpage


\newpage


\newpage


\newpage


\newpage


\newpage


\newpage


\newpage


\end{document}